\documentclass{svjour3}
\smartqed  
\usepackage{graphicx}
\usepackage{subfigure}
\usepackage{booktabs} 
\usepackage{natbib}
\newcommand{\url}[1]{\texttt{#1}}
\newcommand{\urlnote}[1]{\footnote{\url{\texttt{#1}}}}
\newcommand{\forget}[1]{}

\newenvironment{smalltabular}[1]
{\begin{center}
    \begin{small}
        \begin{tabular}{#1}}
{        \end{tabular}
    \end{small}
\end{center}}
\newenvironment{smallertabular}[1]
{\begin{center}
    \begin{footnotesize}
        \begin{tabular}{#1}}
{        \end{tabular}
    \end{footnotesize}
\end{center}}

\newcommand{\thsup}[0]{\textsuperscript{th}}

\newcommand{\xmlopen}[1]{\texttt{<#1>}}

\newcommand{\xmltags}[1]{\xmlopen{#1} tags}

\begin{document}
    \title{
        Building a VO-compliant Radio Astronomical DAta Model for
        Single-dish radio telescopes (RADAMS)
    }
    \titlerunning{Building a VO-compliant Radio Data Model (RADAMS)}
    
    
    \author{
        J.D.~{Santander-Vela}\and
        E.~{Garc\'ia}\and
        S.~{Leon}\and
        V.~{Espigares}\and
        J.E.~{Ruiz}\and
        L.~{Verdes-Montenegro}\and
        E.~{Solano}
    }
    
    \institute{
        J.D. Santander-Vela \at European Southern Observatory, ESO,
        Karl-Schwarzschild-Strasse 2, 85748 Garching,
        Germany. \email{jdedsant@eso.org}
        \and
        J.D. Santander-Vela \and E. {Garc\'ia} \and V. {Espigares} \and 
        J.E. {Ruiz}\and L. {Verdes-Montenegro}
        \at Instituto de Astrof\'isica de Andaluc\'ia, IAA-CSIC,
        Camino Bajo de Hu\'etor 50, 18080 Granada, Spain.
        \and
        S. Leon \at Joint ALMA Observatory/European Southern Observatory, JAO/ESO,
        Av. Alonso C\'ordova 3107, Vitacura,
        Santiago, Chile.
        \and
        E. Solano \at Centro de Astrobiolog\'{\i}a. Departamento de Astrof\'{\i}sica. P.O. 78. 
        E-28691 Villanueva de la Ca\~nada, Madrid, Spain.
    }

    \date{Received: date / Accepted: date}    

    \maketitle

    \begin{abstract}
        The Virtual Observatory (VO) is becoming the de-facto
        standard for astronomical data publication. However, the
        number of radio astronomical archives is still low in
        general, and even lower is the number of radio astronomical
        data available through the VO. In order to facilitate the
        building of new radio astronomical archives, easing at the
        same time their interoperability with VO framework, we have
        developed a VO-compliant data model which provides
        interoperable data semantics for radio data. That model,
        which we call the Radio Astronomical DAta Model for
        Single-dish (RADAMS) has been built using standards of (and
        recommendations from) the International Virtual Observatory
        Alliance (IVOA). This article describes the RADAMS and its
        components, including archived entities and their 
        relationships to VO metadata. We show that by using IVOA 
        principles and concepts, the effort needed for both the 
        development of the archives and their VO compatibility has 
        been lowered, and the joint development of two radio
        astronomical archives have been possible.
        We plan to adapt RADAMS to be able to deal with
        interferometry data in the future.
        
        
    \end{abstract}

    \section{Introduction} 
    \label{sec:introduction}
        
        
        With the advent of digitisation, sharing astronomical data
        has become not only possible, but desirable in order to make
        better use of our community's valuable observing resources,
        and combine data from multiple instruments to enhance
        science, making data exploitation more efficient.
        
         However, as more and more data are made available by
        observatories and other data providers in the form of
        digitally archived data, astronomers face new
        problems. They have to deal with different data access
        procedures, different data formats, and different data
        semantics (how to compare equivalent concepts
        originating from different instruments).
        
         The VO \citep{2001Sci...293.2037S} tries to answer these
        concerns by specifying common data access protocols such as
        the Simple Cone Search \citep{2009scs..rprt.....W} for
        spatially querying catalogues, or the Simple Image Access
        and Simple Spectra Access Protocols
        \citep{2009sia..rprt.....T, 2008ssa..rprt.....T} for image
        and spectral data; data formats such as the XML-based
        VOTable \citep{2009votfdivoav3011O} for tabular data
        interchange, linking to FITS files, and expressing
        observation metadata; and data semantics, by means of
        virtual observatory data models.
        
         These common standards have been possible by the joint work
        of an international standardisation body, the IVOA
        \citep{Hanisch:2003vn}, a federation of national and
        supranational VO groups, which steers and sanctions the
        development of the different parts of the VO infrastructure,
        thanks to its different Working Groups (WGs). In particular,
        the Data Access Layer (DAL) and Data Modelling (DaM) WGs are
        the ones in charge of standardising access protocols and
        data models within the VO.
        
         A data model can be defined \citep[see, for
        instance,][]{Hirschheim:1995qy} as the description of the
        set of entities needed for information storage in a
        particular problem domain to be solved by a software system,
        and specifies both the data being stored, and the
        relationships among them.
        
         Questions
        about the data must be answered by using the relationships 
        encoded in the data model. Considered this way, a uniform, 
        interoperable observation
        data model can be mapped into a uniform set of questions
        which can be answered within the VO.
        
         VO data models affect how data from observatories' archives
        are exposed through VO services, as the way such data are
        stored by the archive might differ from the IVOA standards.
        However, there are particular attributes that need a precise
        description in order for the model to be useful,
        so VO data models need to fix the kind
        of metadata they support: a completely open model, where
        arbitrary metadata can be attached, can be very useful for
        annotations, but it is much more difficult to use for
        querying. And we must have present that, within the VO,
        data models apply not only to the scientific data, but to
        the metadata describing them.
        
         Both within and outside the VO framework there is a long
        history and expertise on building optical, infrared and
        ultraviolet archives \citep[e.g., MAST, see][for a 
        VO perspective]{1999ASPC..172..233C, 2005ASPC..347..208K}.
        However, that expertise is not as pervasive for radio
        astronomy, where only a few of the radio telescopes provide
        an online archive, and very few observations are accessible
        through the VO.
        
         An interesting challenge, then, is building radio
        astronomical data archives both for single-dish and
        interferometry instruments, accessible through the VO, 
        enhancing the multi-wavelength window available through
        the VO.
        But for that, we also need
        to provide an answer to the question on how to express the
        data access, data format, and data semantics of radio
        astronomical data within the framework of the VO, so that
        these datasets can be easily combined with data from
        other wavelengths. This
        article is focused on the data format and data semantics for
        such services, using as much as possible of the existing VO
        infrastructure and standards, and avoiding the reinvention
        of existing processes.
        
         The work presented here was developed within the AMIGA project
        (Analysis of the interstellar Medium of Isolated
        GAlaxies\urlnote{http://amiga.iaa.csic.es}) at the
        Instituto de Astrof\'{\i}sica de Andaluc\'{\i}a (IAA-CSIC)
        with the aim of providing an archival infrastructure for the
        Institut de RadioAstronometrie Millim\'etrique
        (IRAM\urlnote{http://www.iram-institute.org/}), in order
        to lower the barriers for radio astronomical data publishing
        in the VO, and to increase the number of available datasets
        in the radio wavelengths. This was possible because AMIGA
        scientific activities \citep[e.g.][]{2005A&A...436..443V}
        are being supported by a group of software engineers
        specialised in radio archives within the VO framework
        \citep[see ][for a review of AMIGA VO
        activities]{2008arXiv0810.2317R}.
        
        
    
    \section{Building blocks: available IVOA data models} 
    \label{sec:available_ivoa_data_models}
        
        Given the decision to bring VO compatibility to the radio
        astronomical archives to be developed from the beginning, we
        wanted to make as much reuse as possible of existing VO
        standards and best practices.
        
         By 2005, the beginning of the development of the the TAPAS 
        IRAM-30m \citep[see ][]{Leon:2012fk_ea} and DSS-63 archives, 
        and hence of the RADAMS, 
        the IVOA had only
        one Recommendation~\citep[highest degree of endorsement by
        the IVOA; see][]{2009idstdivoav0302H}, the
        VOTable\urlnote{http://www.ivoa.net/Documents/VOTable/}
        \citep{2009votfdivoav3011O}. This provided a common format
        for tabular data interchange, but did not provide specific
        semantics for it.
        
         The Data Model for Observation \citep[ObsDM;\
        ][]{2005dmo..rept.....M} was drafted by the DaM WG to
        provide a framework that could be used to unify all kinds of
        astronomical observations from a discoverability point of view.
        It provides the concept of
        dataset characterisation, that was later developed in depth
        by the Characterisation data model (CharDM), which was
        finally promoted to the Recommendation status
        \citep{2008dmadcrept.....L}.
        
         The DaM WG has also distilled the minimum (core) metadata elements 
        from the ObsDM  
        \citep{2011arXiv1111.1758L}. These ObsDM Core Components
        can be used to expose and discover through the VO
        lists of observations, by specifying most of the CharDM
        metadata for all kinds of astronomical observations in the
        spatial, temporal, spectral, and polarisation axes.
        
         Other IVOA data models presently in the Recommendation
        status, and which are relevant to the description of radio
        astronomical datasets are the Spectral data model
        \citep[SpecDM;\ ][]{2007sdm..ivoa.....M}, and the Space-Time
        Coordinate metadata model\footnote{As of this writing, the
        Simple Spectral Lines data model is in the Request For
        Comments (RFC) phase, and if no negative responses are
        presented, will be promoted the Recommendation status
        \citep{2010ssldmivoav1005O} as soon as revised draft, taking
        into account the comments provided during the RFC, is
        submitted to the IVOA's Executive committee.} \citep[STC;\
        ][]{2007stc..ivoa.....R}.
        
         In addition to data models, the IVOA also provides
        recommendations for standardised data semantics, so that
        equivalent information between data models can be tagged and
        identified as such. The Unified Content
        Descriptors~\citep[UCDs; ][]{2007ucd..ivoa.....P} are atomic
        units which can be used to identify the physical meaning of
        different data columns, or data attributes, with rules for
        combining them to provide more specific meaning, or to match
        them against search terms to find equivalent
        concepts~\citep{Derriere:2004lr}. On the other hand, UTypes
        where defined as an specific attribute which can be attached
        to a VOTable \xmlopen{FIELD} or \xmltags{PARAM}, in order to
        make clear that the values for that field or parameter
        exactly correspond to a particular data model
        \citep[see][section 4.6]{2009votfdivoav3011O}. We will make
        use of these semantic constructs to be able to export
        archive data with the most rich metadata.
        
         By directly using VO data models as the basis for
        astronomical archives, no intermediate layer for translation
        is needed between the storage and VO publication stage, and
        we ensure at the same time that all relevant VO metadata are
        present.
        
         In case of building VO archives on top of existing, non-VO
        oriented archive, the translation layer needs to provide
        some metadata which are not just translated, but in many
        cases need to be extracted or even calculated from either
        the FITS headers, the actual FITS data tables/images, the
        observation logs, or possibly a combination of all of them.
        
         When dealing with radio astronomical specific concepts not
        covered by any VO standard, we made use of the proposals
        from Lamb and Power~\citep{LamPow0310IVOA} to the IVOA, 
        the VO archive of
        the Australia Telescope Compact Array \citep[ATCA;
        ][]{2006PASA...23...25M}, and concepts from the Alma Science
        Data Model \citep[ASDM; ][]{2006ASPC..351..627V}.
        

    \section{Building the RADAMS} 
    \label{sec:the_radams}
        
        While using the VO, we can identify four different
        phases in the typical astronomer's workflow, with
        different data semantics needed during each one:
    
        \begin{itemize}
            \item \emph{Discovery:}
            Datasets available in the VO
            have to be discoverable for them to appear 
            in VO tools. In this phase we need a unified description
            of where in space-time, but also on frequency (possibly
            velocity) coordinates, spatial, temporal, and frequency
            resolution, the VO Registry holds data for existing
            datasets so that they can be easily discovered. The data
            model for STC (albeit in a simplified form most of
            the time), CharDM, Resource metadata (ResDM),
            and the UCD and IVOA thesaurus (IVOAT) are relevant in
            this phase.
            
            \item \emph{Evaluation:}
            Datasets have to be evaluated
            in order to assess their applicability to the kind of
            analysis we might wish to perform; for instance, in
            order to do image mosaicing we need a certain coordinate
            overlap, and in order to do image stacking we need an
            almost complete overlap, and comparable resolutions. The
            main data model involved in this phase is the CharDM,
            but also Curation (to evaluate data based on authority),
			and Provenance (to evaluate data based on observational 
			and environmental restrictions).
            
            \item \emph{Data Access:}
            There is an implicit data
            model in the IVOA data access protocols, the Data Access
            Layer (DAL), which is focused on targets (coordinates with
            tolerances/search radii), and uses several properties
            from the CharDM, such as the Coverage in several axes,
            to allow for narrower searches. This DM is always used
            by all systems.
            
            \item \emph{Transformation:}
            When creating a new
            data set, or transforming an existing one, a new CharDM
            instance needs to be created. If the transformed data
            set is a spectrum, the Spectral data model (SpecDM) is
            needed both for obtaining the complete description of
            the original data and describing the transformed
            product\footnote{Incidentally, there is no existing 
            data model yet for images
            or for more complex data within the VO, beyond what is
            available from FITS conventions.}.  In addition, in
            order to trace the origin of the transformed image we
            would need to use interoperable Provenance information,
            suggested by the ObsDM.
        \end{itemize}
        
        We can see, then, that in order to properly enable VO
        operations for dataset Evaluation and Transformation
        we need to use Provenance metadata. 
		Provenance-related dataset metadata are part of some
		IVOA data models (i.e., the SpecDM DataID), but RADAMS aims
		to provide a description of Provenance applicable to all 
		kinds of radio observations.
		Besides, Curation
        metadata, part of the Discovery process, need to be rich
        enough to be linked with Policy information. Plus, 
        if we want to deliver data beyond FITS files, and allow
        characterisation information to be available in native
        format offline, Packaging is needed. By adding extra,
        radio astronomical specific metadata, we also enhance
        the specification of Sensitivity in the CharDM.
    
        \begin{figure}[tb]
            \centering
                \includegraphics[width=1\columnwidth]
                {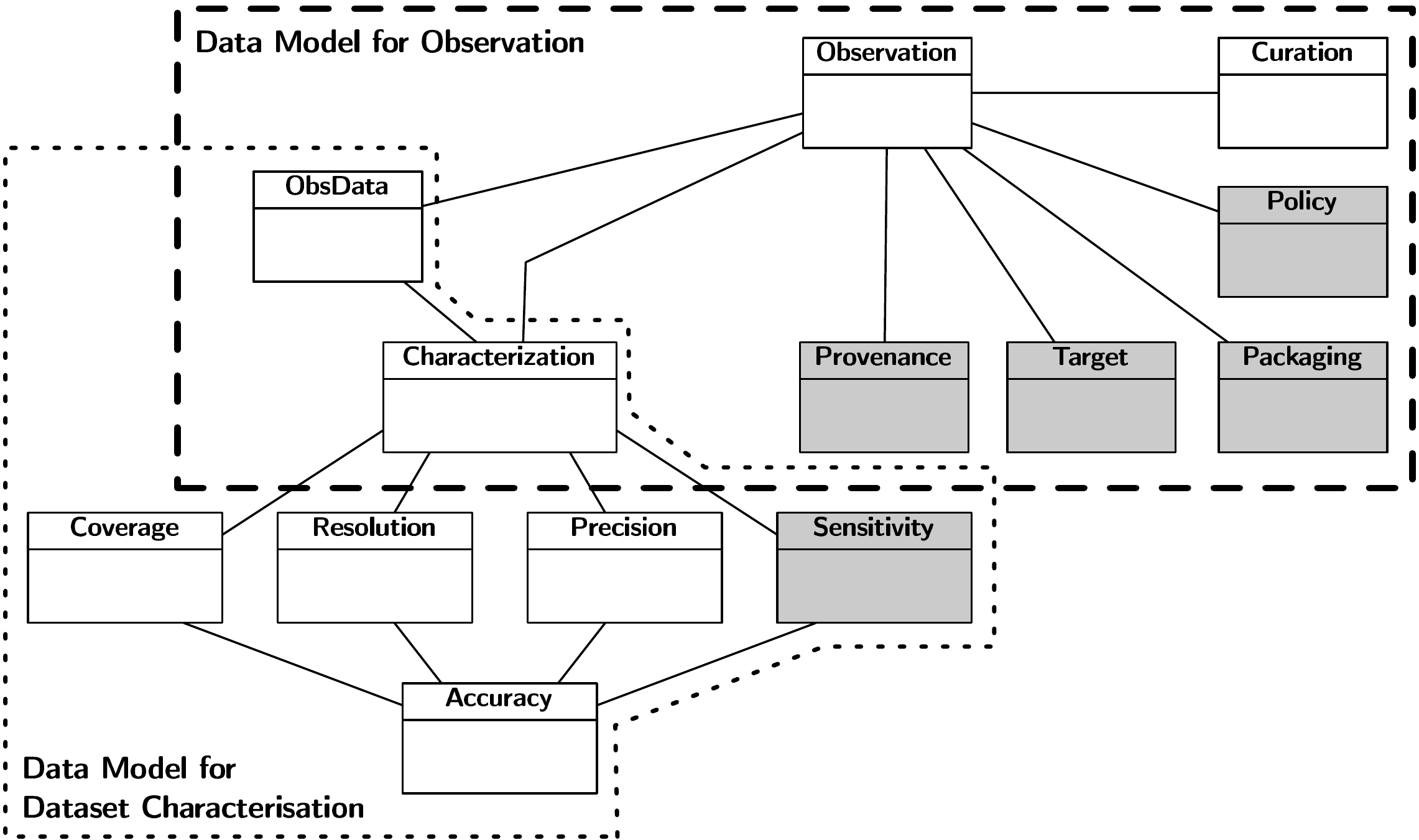}
            \caption{The complete RADAMS data model, showing the
            inspiration of their classes, with classes newly
            specified by the RADAMS in grey.}
            \label{fig:fig_CompleteRobledoDataModelWithOrigins}
        \end{figure}
        
         This shows that the RADAMS is a radio astronomical 
        observation data
        model whose main emphasis is describing single dish radio
        observations (hence, Radio Astronomical DAta Model for
        Single-dish), by using the ObsDM for VO standardised
        metadata. RADAMS implements for the first time
		for radio astronomical observations,
		irrespective of the data product being generated,
		parts of the
        ObsDM such as Curation, Policy,
        Provenance, and Packaging, and defining how to calculate the
        radio astronomy specific parts of the CharDM.
        
         Figure~\ref{fig:fig_CompleteRobledoDataModelWithOrigins}
        shows the main composition of the RADAMS, with the origin of
        the different set of classes, and shows greyed out the
        classes which are defined by the RADAMS for the first time.
        
         RA\-DAMS' defining classes and sub-models are:
        
        \begin{itemize}
            \item \emph{Observation:} Abstract root class for the 
            data model.
            Works as a hub to which both actual observational
            data (ObsData) and metadata are linked
            together.
            
            \item \emph{ObsData:} Represents the actual data being
            described by all the RADAMS' classes.
            
            \item \emph{Target:} Describes the target of the
            observation, providing as much information as available
            for already known targets, and only pointing information
            for not known targets.
            
            \item \emph{Characterisation:} Corresponds to the
            Char\-DM classes, describing where to locate the
            observation in scientific parameter space (spatial
            bounds, temporal bounds, spectral bounds, and
            even observed flux bounds or polarisation), and 
            resolution along those axes.
            
            In practice, a different class exists for each different
            kind of axis, as specialising per axis allows for
            different axes having different metadata attached.
            
            \item \emph{Provenance:} Binds all the information
            regarding how the observation was originated, both from
            technical and scientific point of view.
            
            \item \emph{Curation:} Describes who is responsible for
            maintaining a particular observation, a set of
            observations, or all of the archive. It also
			contains the information identifying
			who originated the observation (i.e., the PI of an
			observation program).
            
            \item \emph{Policy:} Is used to specify the access
            rights for different systems and persons accessing the
            archive. Information from the Curation class is used in 
            order to assess the accessibility of datasets.
            
            \item \emph{Packaging:} Describes the way an
            observation, or a set of observations, are actually
            delivered when the archive is queried. Enables
            component reuse, recursive embedding, and 
            sub-package selection.
        \end{itemize}
        
        RADAMS attributes are assigned their semantic meaning using a
        flat table that relates each attribute to its definition. UCDs
        and UTypes are used to relate RADAMS attributes to common
        semantic definitions in existing IVOA data models. 
        These mappings collectively constitute the RADAMS data model.
        
        All of the metadata classes listed before are implemented by
        the RADAMS. In the following sections we will describe in
        detail only the main contributions of the RADAMS, namely
        Policy, Packaging, and specially observation Provenance.
        The description includes tables describing the 
        specific metadata, semantics, and UCDs assigned to 
        each metadata entity.
        
        The medatata for each class will be shown in Tables 1 to 17,
        and referred to appropriately in the following sections.
        For each table, the RADAMS attribute name is listed together
        with its description, UCD, and where available the 
        corresponding FITS keyword either from the FITS
        standard~\citep{FITSWG:2008ty},
        or IRAM Multi-Beam FITS~\citep{MudPolHat0512Multi-Beam}
        or NRAO extensions~\citep{PreCla0412Device}.
        When a clear candidate for a FITS keyword was not found,
        the table will show
        \texttt{N/D} (Not Defined). For information encoded as
        FITS comments, the \texttt{COMMENT} keyword will be used.

    
    \section{Specifying access policy} 
    \label{sec:policy}
        
        Policy is defined as
        \emph{a course or principle of action a\-dop\-ted or proposed by
        a government, party, business, or 
        individual}~\citep{NOAD_OUP:2005lr}. For the VO,
        policies we are concerned with are \emph{data access
        policies}, and they can be defined as the 
        granting or denial of access to data (or metadata) following 
        some principles
        proposed by data providers\footnote{Other policies of interest
        for an astronomical archive, but outside of the scope of the
        VO, might involve data erasure, replication, or billing.}.
        Those principles take into
        account the agents (people, groups of people,or systems) 
        responsible for data generation or data curation, and the 
        relationship between these agents and the user trying to access 
        the data.
        
         Moreover, those principles (policies) change from
        institution to institution, and also for different datasets
        curated by those institutions. Hence, 
        the way to specify those different policies must be general
        enough, and must take into account
        the different users' relationships with the datasets, and
        the different ways to implement policies, from
        \emph{everything is accessible}, to \emph{PI's eyes only},
        and going through \emph{metadata for everyone, data only for
        PI and CoIs}, among other possibilities\footnote{An example
        of very user oriented policy can be found at ESO, where
        certain datasets are only accessible by users who are
        registered as belonging to an ESO member state.}.
        
         The Policy data model must be able to allow for very
        complex policies to be applied to the data. In the case of a
        VO archive where the data are not instantly available
        as part of an operational workflow, or
        for data which are intended to be public from the moment of
        insertion, Policy becomes simpler\footnote{In the case of
        the Robledo Archive, the policy implemented was the standard
        NRAO policy: 18 months since the end of the observations.},
        as in \emph{everything in the archive is
        available for everyone as soon as it is in the system}.
        
         The proposed Policy data model implements a Role-Based
        Access Control (RBAC) system \citep[see][]
        {Ferraiolo:1992uq, rbac-features-motivations-1995}, in order
        to simplify the administration of access permissions. RBAC
        greatly simplifies the management of access, and at the same
        time provides greater flexibility. Users can be 
        assigned to roles following the logic of their relationship
        with the data to be accessed (usually corresponding to the
        organisational relationships, i.e., whether users are PI,
        CoIs, observers, operators, delegates, etc. with respect to
        the data). When needed, roles can be granted new permissions
        (accessing new kinds of metadata, for instance), or
        permissions revoked, in a way which is independent of the
        logic rules assigning roles to the users.
		
		 One of the most successful RBAC-based data systems is
		the Integrated Rule-Oriented Data System \citep [iRODS;][]
		{10.1007-978-3-540-69389-5-37}, which uses
		rules not just for data access, but also for data storage 
		and deployment.
        
         Role-based policies are more flexible within the astronomy
        domain than user or group related policies: 
        as the actual file or metadata access or modification 
        permissions will be the same for users with the same role,
        systems administrators have less administrative burden.
        Roles only need to be assigned (or calculated) for users;
        the corresponding permissions are already predefined for
        each role.
        
\begin{figure}[tb]
            \begin{center}
                \includegraphics[width=\columnwidth]
                {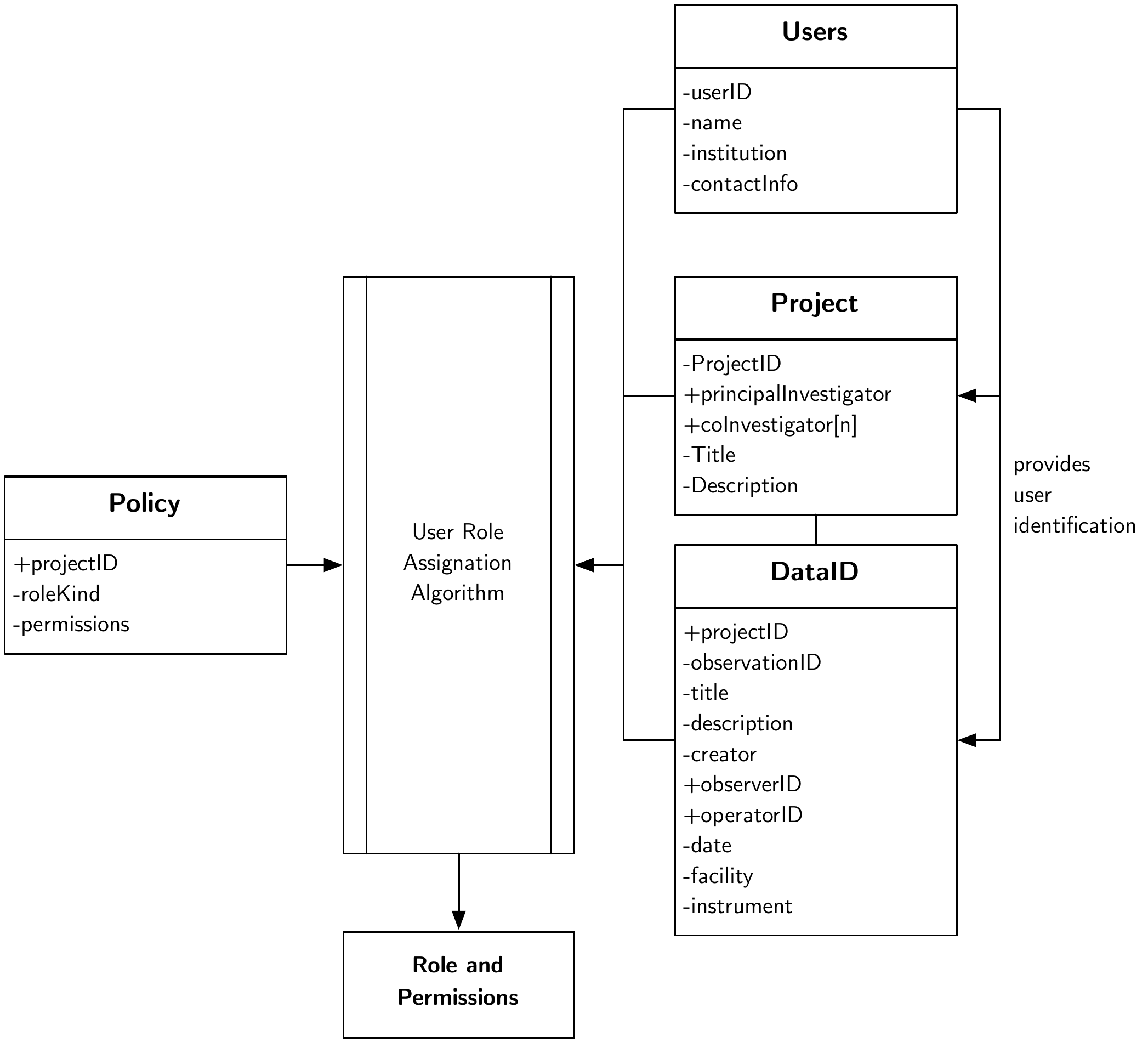}
            \end{center}
            \caption[Policy data model]
                {Policy data model, with role and permissions.}
            \label{figPolicyDataModel}
        \end{figure}

         Figure~\ref{figPolicyDataModel} shows the different classes
        needed to characterise the archive policy in a more generic
        setup.
        
         In RADAMS, roles are chosen from a controlled vocabulary
        (\texttt{prin\-ci\-pal\-In\-ves\-ti\-ga\-tor},
        \texttt{observer}, \texttt{co\-Inves\-tiga\-tor},
        \texttt{observatory\-Staff}, \texttt{none}); roles are
        dynamically derived from user/project relationships, so that
        people not belonging to the observatory, and who have
        nothing to do with the project, would default to
        \texttt{none}.
        The \texttt{none} role ensures that
        there is always at least one role to be matched, with the
        most restrictive access permissions to be implemented.
        
        New roles can be created, with their own permissions
        set, but would only be assigned to a user if logic for
        the user-data-role matching is provided.
        
         Each project in the archive should have, at least, an
        explicit policy of what is allowed for someone with no
        relationship with the Principal Investigator, and people
        with \texttt{prin\-ci\-pal\-In\-ves\-ti\-ga\-tor} roles
        have all access rights to the archive; people with roles
        other than \texttt{prin\-ci\-pal\-In\-ves\-ti\-ga\-tor}
        would fallback to the \texttt{none} role, if the
        permissions for their role are not explicitly declared.
        
\begin{figure}[tb]
        \begin{center}
        \includegraphics[width=0.9\columnwidth]
            {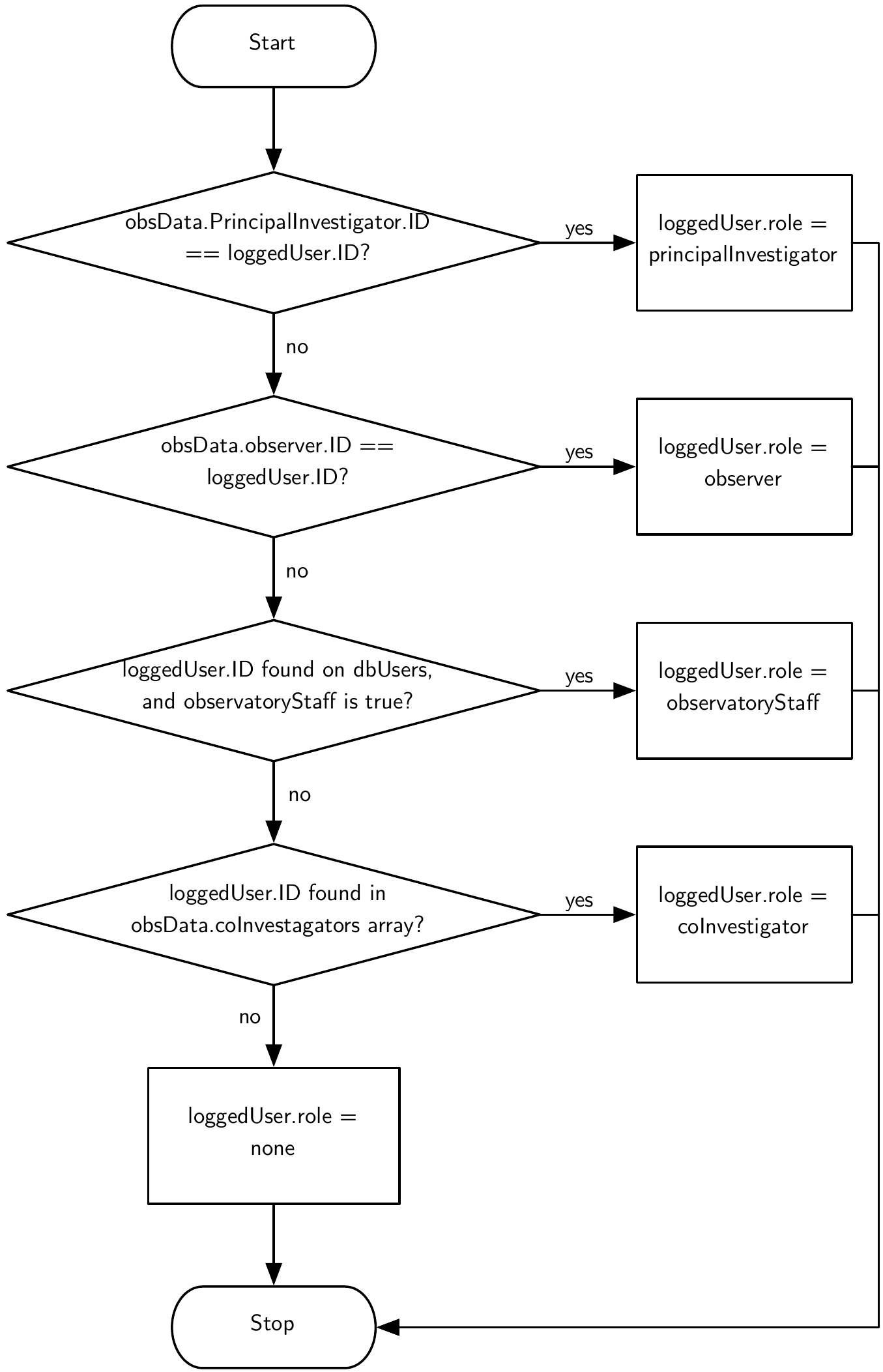}
        \end{center}
        \caption[Role determination algorithm]
            {Flow diagram for the role determination algorithm.}
        \label{figPolicyRoles}
        \end{figure}

         Therefore, we use Policy, Users and ObsData metadata in 
        order to select the corresponding role for the
        agent (a user, or a software agent on their behalf)
        just logged in. Figure~\ref{figPolicyRoles} shows the flow
        diagram for the role selection.
		
		 Although specialised RBAC description and enforcement
		systems such as the eXtensible Access Control Markup 
		Language \citep[XACML;][]{Moses:2005lr}, or the 
		aforementioned iRODS exist, and
		could be used, the RADAMS
		implements a much simpler mechanism, as the number of
		foreseeable roles is limited.
		 
         Policy metadata and attributes are specified in Table
        \ref{tabPolicyMetadata}. We will also need at least a 
        subset
        of Curation attributes for successful Policy
        attribution. These are indicated in Tables 
        \ref{tabPolicyUsersMetadata}, 
        \ref{tabPolicyProjectMetadata},
        and \ref{tabPolicyDataIDMetadata}.
        
        \begin{table*}
        \begin{minipage}{\linewidth}
        \caption[Policy metadata]{Policy metadata.}
        \begin{smallertabular}{p{2.4cm}p{1.5cm}p{3.8cm}p{7.9cm}}
                & \textbf{FITS} & & \\ \textbf{Attribute} & \textbf{Keyword} &
                \textbf{UCD} & \textbf{Description}\\ \midrule  projectID &
                \texttt{PROJID} & \texttt{meta.curation.project;
                meta.id}\footnote{There is no \texttt{meta.curation.project}
                UCD, but we propose the inclusion of one.} & Project
                identifier.\\ \addlinespace roleKind & \texttt{N/D} &
                \texttt{meta.policy.role; meta.code; meta.id}\footnote{There
                are no \texttt{meta.policy.*} UCDs, but we propose, at least,
                the inclusion of the \texttt{meta.policy} atom.} & Role kind
                for which permissions will be provided.\\ \addlinespace permissions &
                \texttt{N/D} & \texttt{meta.policy.permissions;
                meta.code}$^b$ & Permissions provided for roleKind users of
                this project. The particular permissions to be provided are to
                be discussed by IVOA.\\ \addlinespace
        \end{smallertabular}
        \label{tabPolicyMetadata}
        \end{minipage}
        \end{table*}
        
        \begin{table*}
        \begin{minipage}{\linewidth}
        \caption[Policy related Users metadata]
            {Policy related Users metadata.}
        \begin{smallertabular}{p{2.4cm}p{1.5cm}p{3.8cm}p{7.9cm}}
            & \textbf{FITS} & & \\ \textbf{Attribute} &
            \textbf{Keyword} & \textbf{UCD} & \textbf{Description}\\
            \midrule
            
             userID & \texttt{COMMENT} & \texttt{meta.id} & User
            identifier for all user related operations in the
            archive. \\[2pt] name & \texttt{COMMENT} &
            \texttt{meta.name} & Real name of the user. Any known
            user of the archive has to be registered, or be
            anonymous. \\[2pt] institution & \texttt{COMMENT} &
            \texttt{meta.name} & Name of the institution to which
            the user belongs. \\[2pt] contactInfo & \texttt{COMMENT}
            & \texttt{meta.note} & Contact info (probably e-mail)
            for this user. \\[2pt]
        \end{smallertabular}
        \label{tabPolicyUsersMetadata}
        \end{minipage}
        \end{table*}
        
        \begin{table*}
        \begin{minipage}{\linewidth}
        \caption[Policy related Project metadata]
            {Policy related Project metadata.}
        \begin{smallertabular}{p{2.4cm}p{1.5cm}p{3.8cm}p{7.9cm}}
                & \textbf{FITS} & & \\ \textbf{Attribute} & \textbf{Keyword} &
                \textbf{UCD} & \textbf{Description}\\ \midrule  projectID &
                \texttt{PROJID} & \texttt{meta.curation.project;
                meta.id}\footnote{There is no \texttt{meta.curation.project}
                UCD, but we propose the inclusion of one.} & Project
                identifier.\\ \addlinespace principalInvestigator & \texttt{COMMENT} &
                \texttt{meta.id} & User identifier for the
                principalInvestigator of the project. \\ \addlinespace
                coInvestigator[n] & \texttt{COMMENT} & \texttt{meta.id} & User
                identifier for the n\thsup\ co-investigator.\\ \addlinespace
                title & \texttt{COMMENT} & \texttt{meta.curation.project;
                meta.title}$^a$ & Project title.\\ \addlinespace description &
                \texttt{COMMENT} & \texttt{meta.curation.project;
                meta.note}$^a$ & Project description.\\ \addlinespace
        \end{smallertabular}
        \label{tabPolicyProjectMetadata}
        \end{minipage}
        \end{table*}
        
        \begin{table*}
        \begin{minipage}{\linewidth}
        \caption[Policy related DataID metadata]
            {Policy related DataID metadata.}
        \begin{smallertabular}{p{2.4cm}p{1.5cm}p{3.8cm}p{7.9cm}}
                & \textbf{FITS} & & \\ \textbf{Attribute} & \textbf{Keyword} &
                \textbf{UCD} & \textbf{Description}\\ \midrule  projectID &
                \texttt{PROJID} & \texttt{meta.curation.project;
                meta.id}\footnote{There is no \texttt{meta.curation.project}
                UCD, but we propose the inclusion of one.} & Project
                identifier.\\ \addlinespace observationID & \texttt{OBSID} &
                \texttt{obs; meta.dataset; meta.id} & User identifier for the
                principalInvestigator of the project. \\ \addlinespace
                coInvestigator[n] & \texttt{COMMENT} & \texttt{obs; meta.id} &
                User identifier for the n\thsup\ project co-investigator.\\
                \addlinespace title & \texttt{COMMENT} &
                \texttt{meta.curation.project; meta.title}$^a$ & Project
                title.\\ \addlinespace description & \texttt{COMMENT} &
                \texttt{meta.curation.project; meta.note}$^a$ & Project
                description.\\ \addlinespace creatorID & \texttt{AUTHOR} &
                \texttt{meta.curation; meta.id} & User identifier for the
                creator of the data entry.\\ \addlinespace observerID &
                \texttt{OBSERVER} & \texttt{obs.observer; meta.id} & User
                identifier for the person performing the observation generating
                these data.\\ \addlinespace operatorID & \texttt{OBSERVER} &
                \texttt{obs.operator; meta.id}\footnote{We propose the
                inclusion of either \texttt{obs.operator} or
                \texttt{instr.operator} as new UCDs to characterise
                operator-related data. However, \texttt{obs.observer} can be
                used when providing both observer and operator at the same
                time.} & User identifier for the person performing operator
                duties while performing the observation.\\ \addlinespace date &
                \texttt{DATE-OBS} & \texttt{time.obs.start} & Date of
                observation.\\ \addlinespace facility & \texttt{TELESCOP} &
                \texttt{instr.obsty} & Facility (observatory) where the
                telescope/instrument resides in.\\ \addlinespace instrument &
                \texttt{INSTRUME} & \texttt{instr.tel} & Instrument performing
                the observation\footnote{In principle, this metadata item
                should contain an instrument-backend pair, but it might 
                be useful for some future facilities to separate attributes 
                for both.}.\\ \addlinespace
        \end{smallertabular}
        \label{tabPolicyDataIDMetadata}
        \end{minipage}
        \end{table*}

    
    \section{Provenance} 
    \label{sec:provenance}
        
        Provenance metadata provides support for the description of
        how data have been generated, for the purposes of
        attribution, data evaluation, and transformation, and in
        some cases even for data access. Provenance metadata within the
		RADAMS is further classified as:
        
        \begin{itemize}
            \item \emph{Instrumental:} Has to do with the
            instrumental setup, i.e., the configuration of all
            observation elements between the original photon source
            and the photon detection equipment.
            
             \item \emph{Environmental:} Has to do with the elements
            in the path of the photon source which cannot be
            controlled by the instrumental setup, but that
            nonetheless affect the photon collection (i.e., by
            causing turbulence, changing the refraction index, or
            absorbing photons). We will register measurable
            environmental parameters in order to estimate possible
            effects and/or defects in the detections.
            
             \item \emph{Processing:} Once raw data are recorded,
            many different processing steps need to be performed in
            order to provide science-ready data, or as it is
            sometimes said, data are provided with the instrument
            signature removed as much as possible. Processing
            Provenance records the different processes and their
            inputs performed in order to achieve the result being
            offered by the archive.
        \end{itemize}
        
        As stated previously, the medata related to the 
		observation proposal, being arguably part of its 
		provenance, will be kept as part of the
        Curation class.
        
        \subsection{Instrumental provenance} 
        \label{sub:instrumental_provenance}
            
            Instrumental provenance in the RADAMS was initially
            inspired by the IVOA note by \cite{LamPow0310IVOA}. The
            aim was to provide a framework that could be adapted to
            single-dish telescopes, but also to radio
            interferometers, and could be used to specify all
            instrument configuration data using the beam as the unit
            by which we build up feeds, antenna, and global
            telescope configuration.
            
\begin{figure*}[p]
            \begin{center}
            \includegraphics[width=1\textwidth]
            {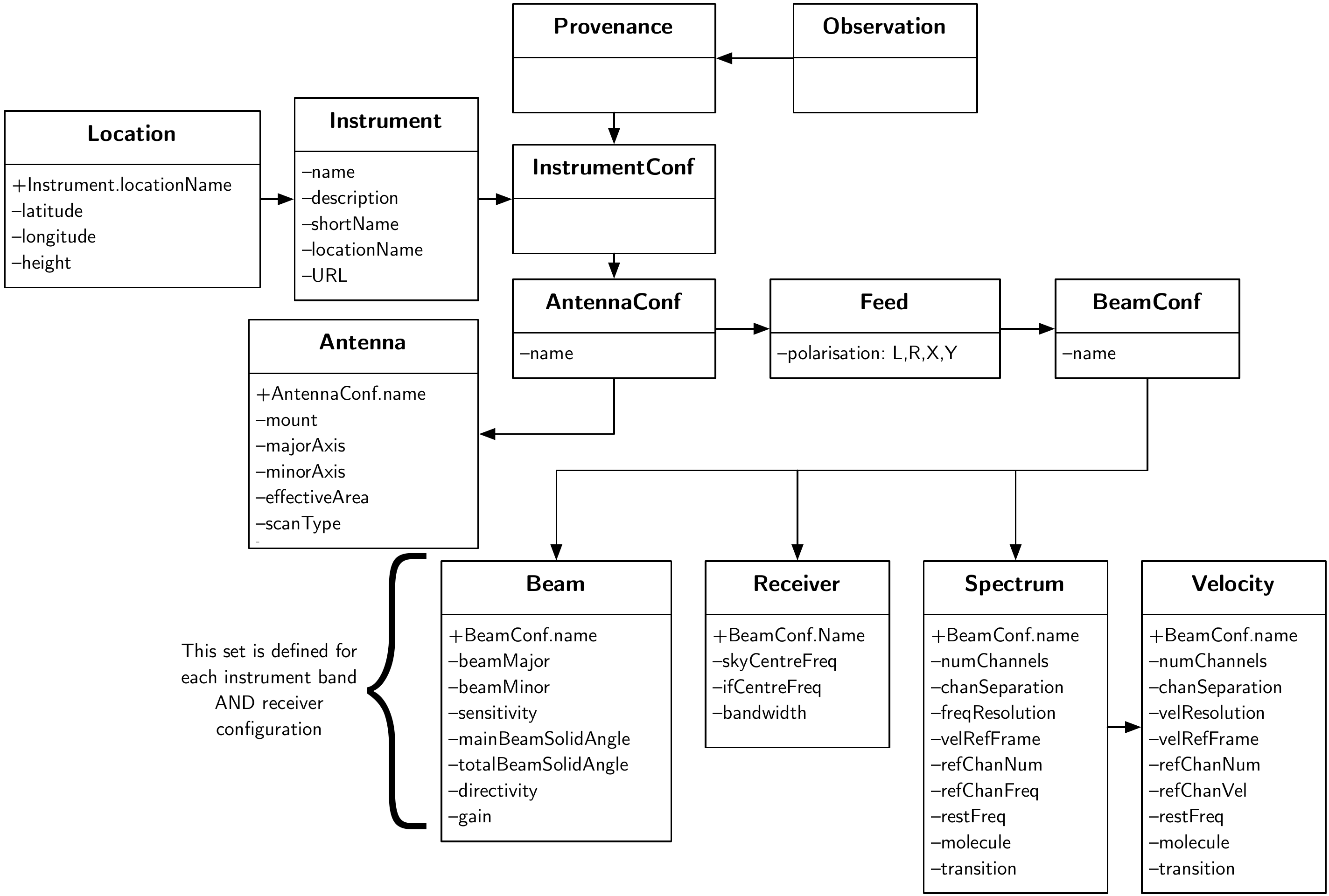}
            \end{center}
            \caption[Provenance.Instrument data model]
            {Provenance.Instrument data model.}
            \label{figProvenanceInstrument}
            \end{figure*}

             Figure \ref{figProvenanceInstrument} shows the classes
            associated with the instrumental configuration for the
            observation.
            
            \begin{itemize}
                \item \emph{InstrumentConf:} Each observation is
                associated to a particular \emph{instrumental configuration},
                which is the result of the particular
                setup of the instrument + antenna + feed
                system. InstrumentConf instances group those
                settings.
                
                 \item \emph{Instrument:} Instances of this class
                specify the instrument-specific part of the configuration,
                as part of the observation setup.
                
                 \item \emph{Instrument.Location:} This is an
                instance of a Location class, used for specifying
                the location for the instrument\footnote{In a 
                generalisation of the RADAMS to radio interferometry,
                aditional locations would have to be applied to each 
                individual Antenna, or an array of locations applied to
                the AntennaConf.}.
                
                 \item \emph{AntennaConf:} In the same way
                InstrumentConf allows the grouping for all the
                instrumental settings, An\-ten\-na\-Conf instances group
                together Antenna and Feed metadata (regarding
                polarisation), plus BeamConf ---another aggregator
                class---. Several AntennaConf instances can provide
                information for different antenna + feed combinations.
                Each possible antenna configuration will be labelled by a
                name. The relationship between
                AntennaConf and Feed instances allows the specification
                of different beams, formed by the combination of
                different feeds. We can use this association
                for a single-dish, multiple-feed configuration where
                each feed went to a different receiver. For single-dish,
                single-feed configurations, there is only one beam 
                associated to a given receiver. 
                
                 \item \emph{Antenna:} Instances of this class
                specify the general properties for each given
                antenna. We will also use these instances to specify
                the type of scan being performed on the source from
                a controlled vocabulary.
                
                 \item \emph{Feed:} Instances of this class ---one
                or more per An\-ten\-na\-Conf--- specify each of the feed
                horns used for the observation, and their
                corresponding polarisation from a controlled
                vocabulary: \texttt{L}, \texttt{R}, \texttt{X},
                \texttt{Y}. We cannot make use of the Stokes
                polarisation parameters, as they cannot be directly
                measured via the feed configuration: instead, they
                have to derived by means of data processing steps.
                
                 \item \emph{BeamConf:} This class is used to group
                multiple beam+receiver pairs for a given feed.
                
                 \item \emph{Beam:} Metadata for this class specify
                the actual beam for the telescope, associated to a
                given spectral band.
                
                 \item \emph{Receiver:} These metadata are used to
                describe the most relevant properties of the
                receiver, such as receiver type, intermediate
                frequency ---in the case of heterodyne stages---, et
                cetera.
                
                 \item \emph{Spectrum:} In case of spectroscopic
                observations, the spectral analyser that has been
                used is specified by instances of this class.
                
                 \item \emph{Velocities:} This class mirrors the
                Spectrum class, and is preferred for those cases
                where velocities are used, instead of frequency.
            \end{itemize}
            
            The detailed Instrumental provenance metadata re\-corded
            within the RA\-DAMS is compiled in Tables 
            \ref{tabProvenanceInstrument},
            \ref{tabProvenanceInstrLocation}, 
            \ref{tabProvenanceInstrAntenna},
            \ref{tabProvenanceInstrFeed},
            \ref{tabProvenanceInstrBeam},
            and \ref{tabProvenanceInstrReceiver}.
            In addition, the description of spectral features
            (also to be reused by the CharDM in the Spectral axis)
            is provided in Table \ref{tabProvenanceInstrSpectrum},
            with a mirror for velocity-based metadata in Table
            \ref{tabProvenanceInstrVelocity}.
			
			 The definition of each element of the \texttt{scanType}
			vocabulary (Table \ref{tabProvenanceInstrAntenna}) can
			be found in Table \ref{tabScanTypes}.
            
             The vocabulary in use for Table
            \ref{tabProvenanceInstrSpectrum} reflects the current
            status of radio astronomical observations, where only a
            line is selected as the target for the observation.
            However, broader spectrometers, such as those available
            with EMIR, or in the future with ALMA, might ask for a
            list of available lines, instead of specifying a single
            one.
            
            \begin{table*}
            \caption[Provenance instrument metadata]
            {Provenance instrument metadata.}
            \begin{smallertabular}{p{2.4cm}p{1.5cm}p{3.8cm}p{7.9cm}}
                & \textbf{FITS} & & \\ \textbf{Attribute} &
                \textbf{Keyword} & \textbf{UCD} & \textbf{Description}\\
                \midrule name & \texttt{INSTRUME} & \texttt{meta.id;
                instr; meta.main} & Instrument name.\\[2pt]
                description & \texttt{N/D} & \texttt{meta.note;
                meta.main} & Instrument description.\\[2pt]
                shortName & \texttt{N/D} & \texttt{instr;
                meta.id} & Short name for the instrument.\\[2pt]
                locationName & \texttt{N/D} & \texttt{instr; pos;
                meta.id} & Instrument site identification.\\[2pt]
                URL & \texttt{COMMENT} & \texttt{instr;
                meta.ref.url} & URL for the instrument (website for
                the instrument, documentation, or any other type of
                instrument description).\\[2pt]
            \end{smallertabular}
            \label{tabProvenanceInstrument}
            \end{table*}

            \begin{table*}
            \caption[Instrument location metadata]
            {Instrument location metadata.}
            \begin{smallertabular}{p{2.4cm}p{1.5cm}p{3.8cm}p{7.9cm}}
                & \textbf{FITS} & & \\ \textbf{Attribute} &
                \textbf{Keyword} & \textbf{UCD} & \textbf{Description}\\
                \midrule locationName & \texttt{N/D} &
                \texttt{instr; pos; meta.id} & Name of a particular
                instrument location.\\[2pt] latitude &
                \texttt{SITELAT} & \texttt{instr; pos.earth.lat} &
                Instrument location latitude.\\[2pt] longitude &
                \texttt{SITELONG} & \texttt{instr; pos.earth.lon} &
                Instrument location longitude.\\[2pt] altitude &
                \texttt{SITEELEV} & \texttt{instr;
                pos.earth.altitude} & Instrument location
                altitude.\\[2pt]
            \end{smallertabular}
            \label{tabProvenanceInstrLocation}
            \end{table*}

            \begin{table*}
            \caption[Antenna configuration metadata]
            {Antenna configuration metadata.}
            \begin{smallertabular}{p{2.4cm}p{1.5cm}p{3.8cm}p{7.9cm}}
                & \textbf{FITS} & & \\ \textbf{Attribute} &
                \textbf{Keyword} & \textbf{UCD} & \textbf{Description}\\
                \midrule name & \texttt{N/D} &
                \texttt{instr.telescope; meta.title; meta.id} & Name
                of the particular antenna.\\[2pt] scanType &
                \texttt{N/D} & \texttt{instr.setup; meta.code} &
                Type of scan being performed by this antenna, from a
                limited vocabulary 
                \citep{LamPow0310IVOA}: \texttt{beam\-Switch},
                \texttt{cal}, \texttt{cross},
                \texttt{dop\-pler\-Track}, \texttt{dwell},
                \texttt{fo\-cus}, \texttt{fre\-quen\-cy\-Switch},
                \texttt{hol\-og\-ra\-phy}, \texttt{mo\-sa\-ic},
                \texttt{on\-Off}, \texttt{on\-The\-Fly},
                \texttt{point}, \texttt{pos\-i\-tion\-Switch},
                \texttt{pul\-sar}, \texttt{ras\-ter},
                \texttt{sky\-Dip}, \texttt{tiedArray},
                \texttt{track}, \texttt{wob\-bler\-Switch}.\\[2pt]
                mount & \texttt{N/D} & \texttt{meta.note} & Mount
                type for the telescope from a limited vocabulary:
                \texttt{azimuthal}, \texttt{e\-qua\-to\-ri\-al},
                \texttt{alt\-az\-i\-muth\-al}, \texttt{dobson},
                \texttt{german e\-qua\-to\-ri\-al}.\\[2pt] majorAxis
                & \texttt{N/D} & \texttt{instr;
                phys.size.smajAxis} & Major axis dimensions.\\[2pt]
                minorAxis & \texttt{N/D} & \texttt{inst;
                phys.size.sminAxis} & Minor axis dimensions.\\[2pt]
                effectiveArea & \texttt{N/D} & \texttt{instr;
                phys.area} & Effective instrument area.\\[2pt]
            \end{smallertabular}
            \label{tabProvenanceInstrAntenna}
            \end{table*}
			
			\begin{table*}
			    \caption[Valid \texttt{scanType} attribute values]{
			        Meaning of the different valid values for
			        the \texttt{scanType} attribute in Table 
					\ref{tabProvenanceInstrAntenna}.
			    }
			    \begin{smalltabular}{rp{8.5cm}}
			        \textbf{scanType} & \textbf{description}
			        \\ \midrule 
        
					\texttt{beamSwitch}      & 
					Scan where the change of reference position is
					accomplished by means of changes in the beam optical
					path.
					\\[2pt]
					\texttt{cal}             & 
					Power or gain calibration scan. \\[2pt]
					\texttt{cross}           & 
					Pointing scan across azimuth and elevation. \\[2pt]
					\texttt{dopplerTrack}    & 
					Tracking with doppler correction for the
					channels being automatically applied.
					Compare with\texttt{track}.\\[2pt]
					\texttt{dwell}           & 
					A tracking scan for which the duration of the actual 
					observation time is prescribed, apart from any time 
					needed to reach the source.\\[2pt]
					\texttt{focus}           & 
					Scan used to focus the system. \\[2pt]
					\texttt{frequencySwitch} & 
					On-off scan performed by means of frequency 
					switching. \\[2pt]
					\texttt{holography}      & 
					Surface precision scanning by means of raster-scanning
					the beam through a well-known point source. \\[2pt]
					\texttt{mosaic}          & 
					Interferometry-specific scan type, for creating
					data cube mosaics. \\[2pt]
					\texttt{onOff}           & 
					On-off scan, without specifying the switching
					technique. \\[2pt]
					\texttt{onTheFly}        & 
					Mapping scan performed while the antenna follows a 
					pre-defined path, with some periodical movement to a 					 reference (off) position. \\[2pt]
					\texttt{point}           & 
					Power or spectral observation on a fixed antenna
					position. \\[2pt]
					\texttt{positionSwitch}  & 
					On-off scan performed by means of changing the 
					antenna reference position. \\[2pt]
					\texttt{pulsar}          & 
					Pulsar-tuned tracking scan.\\[2pt]
					\texttt{raster}          & 
					Power or spectral observations along a raster line. \\[2pt]
					\texttt{skyDip}          & 
					Sky-dip, or tipping of the antenna to get a $T_A$
					or $T_{sys}$ profile over different elevations.\\[2pt]
					\texttt{tiedArray}       & 
					Interferometry-specific scan type, where the complete
					array produces phased data to VLBI observations. 
					 \\[2pt]
					\texttt{track}           & 
					Power or spectral observation while the antenna tracks a
					particular celestial
					point, or follows a solar system body. \\[2pt]
					\texttt{wobblerSwitch}   & 
					Similar to \texttt{beamSwitch}, but using a \emph{wobbler}
					as secondary mirror.
					 \\[2pt]
        
					\end{smalltabular}
			        \label{tabScanTypes}
			\end{table*}

            \begin{table*}
            \caption[Feed configuration metadata]
                {Feed configuration metadata.}
            \begin{smallertabular}{p{2.4cm}p{1.5cm}p{3.8cm}p{7.9cm}}
                & \textbf{FITS} & & \\ \textbf{Attribute} &
                \textbf{Keyword} & \textbf{UCD} & \textbf{Description}\\
                \midrule polarisation & \texttt{POLTY} &
                \texttt{phys.polarization; meta.code} &
                Polarisation value from a controlled vocabulary:
                \texttt{L}, \texttt{R}, \texttt{X}, \texttt{Y}\\[2pt]
            \end{smallertabular}
            \label{tabProvenanceInstrFeed}
            \end{table*}

            \begin{table*}
            \begin{minipage}{\linewidth}
            \caption[Beam configuration metadata]{Beam configuration metadata.}
            \begin{smallertabular}{p{2.4cm}p{1.5cm}p{3.8cm}p{7.9cm}}
                        & \textbf{FITS} & & \\ \textbf{Attribute} &
                        \textbf{Keyword} & \textbf{UCD} & \textbf{Description}\\
                        \hline beamMajor & \texttt{BMAJ/HPBW} &
                        \texttt{instr.beam; phys.size.smajAxis} & Major axis HPBW
                        of the main lobe of the beam.\\ \addlinespace beamMinor &
                        \texttt{BMIN/HPBW} & \texttt{instr.beam;
                        phys.size.sminAxis} & Minor axis HPBW of the main lobe of
                        the beam.\\ \addlinespace sensitivity & \texttt{BEAMEFF} &
                        \texttt{instr.beam; instr.sensitivity} & Beam average
                        sensitivity.\\ \addlinespace mainBeamSolidAngle &
                        \texttt{N/D} & \texttt{instr.beam; pos.posAng;
                        meta.main} & Main lobe’s beam solid angle.\\ \addlinespace
                        totalBeamSolidAngle & \texttt{N/D} &
                        \texttt{instr.beam; pos.posAng; stat.max} & Total beam
                        solid angle, including secondary lobes.\\ \addlinespace
                        directivity & \texttt{N/D} & \texttt{instr.beam;
                        instr.setup; arith.factor} & Directivity percentage.\\
                        \addlinespace gain & \texttt{ANTGAIN} & \texttt{instr.beam;
                        instr.setup; arith.factor} & Beam gain\footnote{We still
                        have to clarify if the gain attribute is related to the
                        directivity concept or not, and if it is related with the
                        receiving stages or not.}.\\ \addlinespace
            \end{smallertabular}
            \label{tabProvenanceInstrBeam}
            \end{minipage}
            \end{table*}

            \begin{table*}
            \caption[Receiver metadata]{Receiver metadata.}
            \begin{smallertabular}{p{2.4cm}p{1.5cm}p{3.8cm}p{7.9cm}}
                & \textbf{FITS} & & \\ \textbf{Attribute} &
                \textbf{Keyword} & \textbf{UCD} & \textbf{Description}\\
                \midrule type & \texttt{BACKEND} &
                \texttt{instr.setup; meta.note} & Receiver type
                (HEMT, Bolo\-meter, SIS, et cetera).\\[2pt]
                skyCentreFreq & \texttt{N/D} & \texttt{src;
                em.radio; em.freq} & Antenna tuning
                frequency.\\[2pt] ifCentreFreq & \texttt{N/D} &
                \texttt{instr.setup; em.freq} & Heterodyne receiver
                intermediate frequency (or list of
                frequencies).\\[2pt] bandwidth & \texttt{BANDWID} &
                \texttt{instr.bandwidth} & Filter-bank total
                bandwidth.\\[2pt]
            \end{smallertabular}
            \label{tabProvenanceInstrReceiver}
            \end{table*}

            \begin{table*}
            \begin{minipage}{\linewidth}
            \caption[Spectrum metadata]
            {
                Spectrum metadata. It might be necessary to change the
                \texttt{MOLECULE} and \texttt{TRANSITI} keywords to
                \texttt{LINE}, for better CLASS compatibility.
            }
            \begin{smallertabular}{p{2.4cm}p{1.5cm}p{3.8cm}p{7.9cm}}
                        & \textbf{FITS} & & \\ \textbf{Attribute} &
                        \textbf{Keyword} & \textbf{UCD} & \textbf{Description}\\
                        \hline numChannels & \texttt{NAXISn} &
                        \texttt{spect; em.freq; meta.number} & Number of spectral
                        channels.\\ \addlinespace chanSeparation\footnote{There is a
                        certain redundancy between the
                        Provenance.Spectrum.chanSeparation attribute and the
                        Coverage.Spectral.Resolution attributes.} &
                        \texttt{N/D} & \texttt{spect; em.freq} & Mean channel
                        separation (in frequency units), or channel frequency
                        separation array.\\ \addlinespace freqResolution &
                        \texttt{FREQRES} & \texttt{spect.resolution; em.freq} &
                        Frequency resolution.\\ \addlinespace velRefFrame &
                        \texttt{SPECSYS} & \texttt{spect; phys.veloc; pos.frame;
                        meta.id} & Identification of the reference system used for the
                        velocity.\\ \addlinespace refChanNum &
                        \texttt{N/D} & \texttt{spect; em.freq; meta.number;
                        meta.ref} & Spectral reference channel.\\ \addlinespace
                        refChanFreq & \texttt{OBSFREQ } & \texttt{spect; em.freq;
                        meta.code; meta.ref} & Spectral reference frequency
                        (observed frequency).\\ \addlinespace restFreq & \texttt{FREQn}
                        or \texttt{RESTFREQ} & \texttt{spect.line; em.freq} &
                        Observed spectral line rest frequency.\\ \addlinespace molecule
                        & \texttt{MOLECULE}\footnote{It might be necessary to
                        change the \texttt{MOLECULE} keyword by \texttt{LINE},
                        for better CLASS compatibility.} & \texttt{spect;
                        phys.mol; meta.id} & Molecule name.\\ \addlinespace transition &
                        \texttt{TRANSITI}\footnote{It might be necessary to
                        change the \texttt{TRANSITI} keyword to \texttt{LINE},
                        for better CLASS compatibility.} & \texttt{spect;
                        phys.atmol.transition; meta.id} & Transition.\\ \addlinespace
            \end{smallertabular}
            \label{tabProvenanceInstrSpectrum}
            \end{minipage}
            \end{table*}

            \begin{table*}
            \begin{minipage}{\linewidth}
            \caption[Velocity metadata]{Velocity metadata.}
            \begin{smallertabular}{p{2.4cm}p{1.5cm}p{3.8cm}p{8.9cm}}
                    & \textbf{FITS} & & \\ \textbf{Attribute} & \textbf{Keyword} &
                    \textbf{UCD} & \textbf{Description}\\ \midrule  numChannels
                    & \texttt{N/D} & \texttt{spect; phys.veloc; meta.number} &
                    Number of velocity channels.\\ \addlinespace
                    chanSeparation\footnote{There is a certain redundancy between
                    the Provenance.Velocity.chanSeparation attribute and the
                    Coverage.Spectral.Resolution attributes.} & \texttt{N/D} &
                    \texttt{spect; phys.veloc} & Mean channel separation (in
                    velocity units), or velocity channel separation array.\\ \addlinespace
                    velResolution & \texttt{N/D} & \texttt{spect.resolution;
                    phys.veloc} & Velocity resolution.\\ \addlinespace velRefFrame &
                    \texttt{N/D} & \texttt{spect; phys.veloc; pos.frame;
                    meta.id} & Identification of the reference system used for the
                    velocity.\\ \addlinespace refChanNum & \texttt{N/D} &
                    \texttt{spect; phys.veloc; meta.number; meta.ref} & Velocity
                    reference channel.\\ \addlinespace refChanVel & \texttt{N/D} &
                    \texttt{spect; phys.veloc; meta.code; meta.ref} & Frequency for
                    the velocity reference channel.\\ \addlinespace restFreq &
                    \texttt{RESTFREQ} & \texttt{spect.line; em.freq} & Observed
                    spectral line rest frequency.\\ \addlinespace molecule &
                    \texttt{MOLECULE}\footnote{It might be necessary to change the
                    \texttt{MOLECULE} keyword to \texttt{LINE}, for better CLASS
                    compatibility.} & \texttt{spect; phys.mol; meta.id} & Molecule
                    name.\\ \addlinespace transition & \texttt{TRANSITI}\footnote{It might
                    be necessary to change the \texttt{TRANSITI} keyword to
                    \texttt{LINE}, for better CLASS compatibility.} &
                    \texttt{spect; phys.atmol.transition; meta.id} & Transition.\\
                    \addlinespace
            \end{smallertabular}
            \label{tabProvenanceInstrVelocity}
            \end{minipage}
            \end{table*}

        
        \subsection{Environmental provenance} 
        \label{sub:environmental_provenance}
            
\begin{figure}[tbp]
            \begin{center}
            \includegraphics[width=0.9\columnwidth]
                {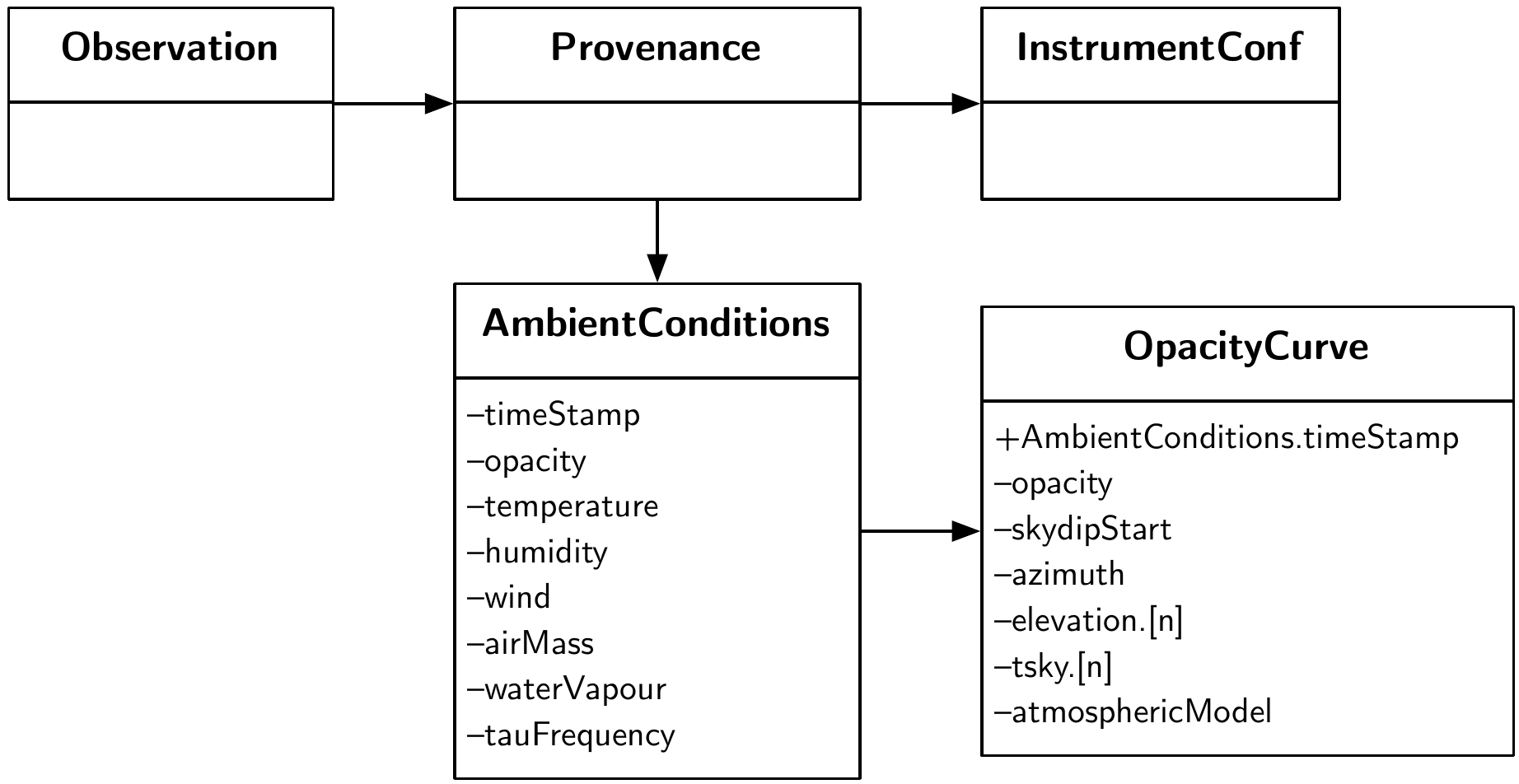}
            \end{center}
            \caption[Provenance.AmbientConditions data model]
                {Provenance.AmbientConditions data model.}
            \label{figProvenanceAmbient}
            \end{figure}

            Environmental provenance is the part of the provenance
            dealing with ambient conditions, and as such the main
            class is called Provenance.AmbientConditions: it
            encompasses all metadata needed to specify weather
            conditions, air mass, opacity, et cetera. Figure
            \ref{figProvenanceAmbient} shows the corresponding
            classes and their relationships.
            
            \begin{itemize}
                \item \emph{AmbientConditions:} Holds all
                metadata related with weather conditions for the
                observation, such as humidity, wind speed,
                opacity at zenith, et cetera.
                
                 \item \emph{OpacityCurve:} Includes the opacity
                curve (linked as a VOTable file) associated to
                the observing term where the data were observed
                (which we will derive from the nearest two
                skydip scans performed before and after the
                observation). We propose the inclusion of an
                array of \texttt{[elevation, Tsky]} pairs,
                together with the azimuth and the starting time
                of the skydip. Calculation of the opacity curve
                is different for bolometric or heterodyne
                observations. It is also necessary to include
                information on the atmospheric model and/or
                software used for opacity fitting. For instance,
                the atmospheric model used by MOPSIC and MIRA,
                the data reduction packages at the IRAM~30m
                antenna, is the \emph{Atmospheric Transmission
                at Microwaves}~\citep[ATM;
                ][]{ParCerSer01Atmospheric}.
            \end{itemize}

            \begin{table*}
            \begin{minipage}{\linewidth}
            \caption[AmbientConditions metadata]
                {AmbientConditions metadata.}
            \begin{smallertabular}{p{2.4cm}p{3.0cm}p{3.8cm}p{5.8cm}}
                & \textbf{FITS} & & \\ \textbf{Attribute} &
                \textbf{Keyword} & \textbf{UCD} &
                \textbf{itemize}\\ \midrule timeStamp &
                \texttt{DATE-OBS} & \texttt{time.obs.start} &
                Moment at which the weather snapshot is
                taken.\\[2pt] opacity & \texttt{TAUZEN} &
                \texttt{phys.absorption.coeff} & Opacity at
                zenith estimated at the observation
                frequency.\\[2pt] airMass & \texttt{AIRMASS} &
                \texttt{obs.airMass} & Air mass at zenith at the
                observing site.\\[2pt] temperature &
                \texttt{TAMBIENT} & \texttt{phys.temperature} &
                Ambient temperature.\\[2pt] humidity &
                \texttt{HUMIDITY} & \texttt{obs.atmos;
                phys.columnDensity} & Ambient humidity.\\[2pt]
                waterVapour & \verb+TAU_WPATH_RD<freq>+\footnote{This 
                keyword is defined within the IRAM MB-FITS bidimensional table. 
                The first column corresponds to \texttt{waterVapour}, 
                the second to \texttt{tauFrequency}.} &
                \texttt{obs.atmos; phys.pressure} & Equivalent
                pressure of the water vapour column.\\[2pt]
                tauFrequency & \verb+TAU_WPATH_RD<freq>+${}^{a}$ &
                \texttt{obs.atmos; em.freq} & Tau radiometer
                frequency.\\[2pt] wind & \texttt{WINDSPEE} &
                \texttt{obs.atmos; phys.veloc} & Wind
                speed.\\[2pt]
            \end{smallertabular}
            \label{tabProvenanceAmbientConditions}
            \end{minipage}
            \end{table*}
            
            \begin{table*}
            \caption[Opacity metadata]{Opacity metadata.}
            \begin{smallertabular}{p{2.4cm}p{1.5cm}p{3.8cm}p{7.9cm}}
                & \textbf{FITS} & & \\ \textbf{Attribute} &
                \textbf{Keyword} & \textbf{UCD} &
                \textbf{itemize}\\ \midrule opacity &
                \texttt{TAUZEN} & \texttt{phys.absorption.coeff}
                & Opacity at zenith estimated at the observation
                frequency.\\[2pt] skydipStart &
                \texttt{DATE-OBS} & \texttt{time.obs.start} &
                Skydip starting time.\\[2pt] azimuth &
                \texttt{AZIMUTH} & \texttt{pos.az} & Skydip
                azimuth.\\[2pt] elevation[n] & \texttt{ELEVATIO}
                & \texttt{pos.el} & Skydip scan elevation.\\[2pt] 
                tsky[n] & \texttt{N/D} &
                \texttt{instr.skyTemp} & Sky temp at
                $\mathrm{n^{th}}$ skydip.\\[2pt] atmosModel &
                \texttt{N/D} & \texttt{meta.modelled;
                obs.atmos; meta.id} & Atmospheric model
                identification.\\[2pt]
            \end{smallertabular}
            \label{tabProvenanceAmbientOpacity}
            \end{table*}
            
            By providing a unified Ambient Provenance model, we
            allow for querying global ambient variables across
            the VO, which can allow for better selection of
            observations, but also for establishing a
            multidimensional quality metric that takes into
            account weather, and also CharDM resolution, and
            that can even be used for weather science and
            forecast, including applications such as
            weather-prediction assisted scheduling
            \citep{Alvarez:2010fk}, but using data across
            observatories.
            
            RADAMS' ambient metadata are listed in Tables
            \ref{tabProvenanceAmbientConditions} and
            \ref{tabProvenanceAmbientOpacity}.
        
        \subsection{Processing provenance} 
        \label{sub:processing_provenance}
            
            The Provenance.Processing class enables the
            specification of processing processes applied to the
            data before archival, including some processes
            necessary for the actual observation, such as the
            determination of the background signal via
            \texttt{frequencySwitching} or
            \texttt{position\-Switching} for background/source
            data comparison.
            
\begin{figure}[tbp]
            \begin{center}
            \includegraphics[width=0.9\columnwidth]
                {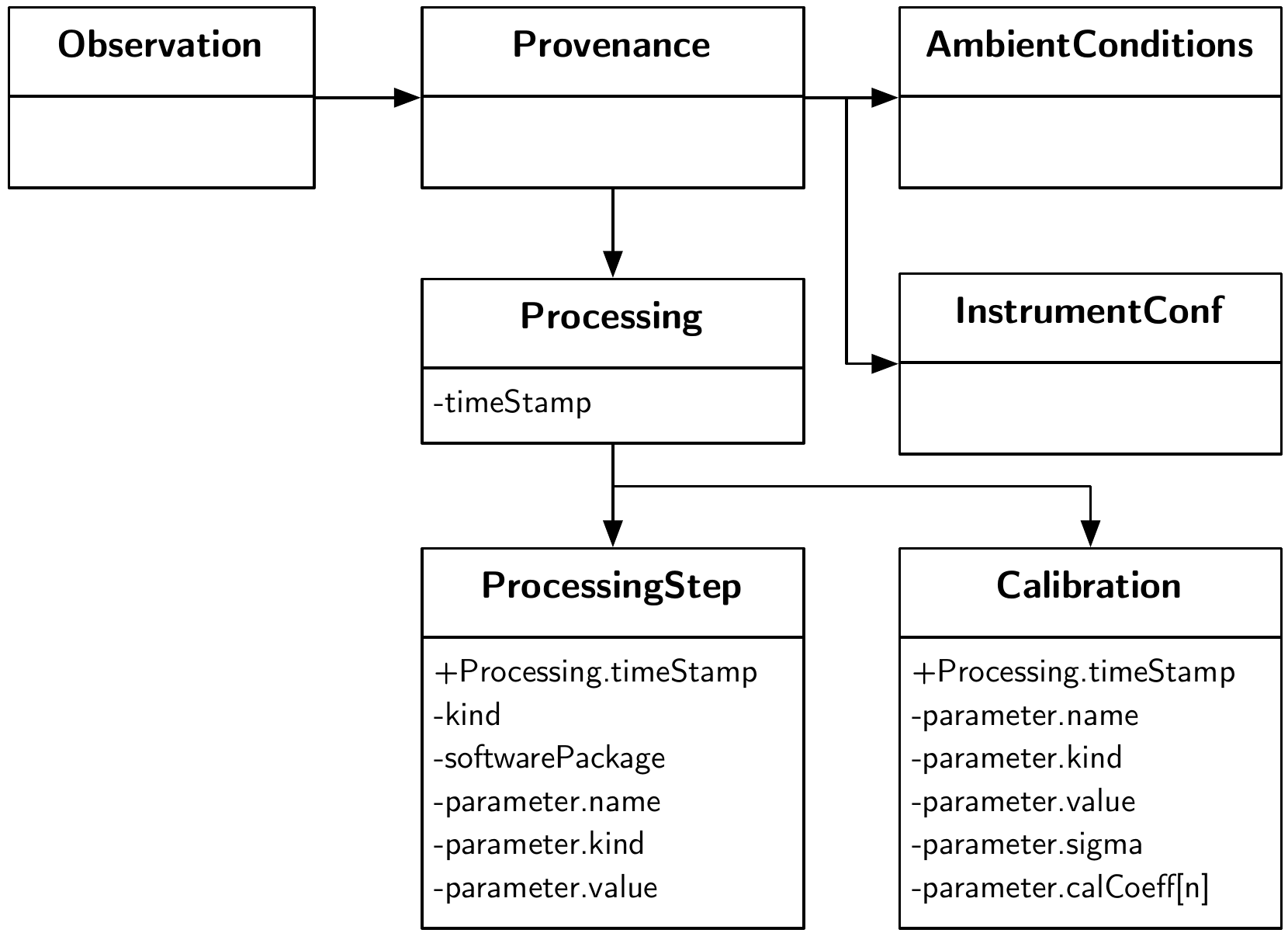}
            \end{center}
            \caption[Provenance.Processing data model]
                {Provenance.Processing data model.}
            \label{figProvenanceProcessing}
            \end{figure}

             The RADAMS will make use of just two classes,
            Processing and Calibration ---this is a subclass of
            proc\-ess\-ing---. Order is relevant, and it should be
            possible to reconstruct the pipeline by the ordering
            of Processing and/or Calibration instances. Figure
            \ref{figProvenanceProcessing} shows these classes.
            
            \begin{itemize}
                \item \emph{Processing:} It holds information
                specifying the type of processing applied to
                data before archival. This includes
                pseudo-observational techniques such as position
                switching or frequency switching, as well as the
                type of data averaging, data weighting, et
                cetera. Table \ref{tabProvenanceProcessingStep}
                provides minimal initial metadata, using arrays
                of parameter keywords for extensibility at the
                expense of complexity.
                
                \item \emph{Calibration:} Is a subclass of
                Processing, where the type of processing is
                \texttt{data\-Calibration}. This class provides
                additional attributes to specify the type of
                calibration, and the axes to where this
                calibration will be applied. Table
                \ref{tabProvenanceCalibration} provides minimal
                initial metadata, using arrays of parameter
                keywords for extensibility at the expense of
                complexity.
            \end{itemize}
            
            We still have to develop a calibration and/or
            pointing model; maybe based upon
            IRAM-Multi-Beam-FITS, or GBT FITS calibration
            tables.
            
            \begin{table*}
            \begin{minipage}{\linewidth}
            \caption[Processing Step]{ProcessingStep metadata.}
            \begin{smallertabular}{p{2.4cm}p{1.5cm}p{3.8cm}p{7.9cm}}
                    & \textbf{FITS} & & \\ \textbf{Attribute} & \textbf{Keyword} &
                    \textbf{UCD} & \textbf{Description}\\ \midrule  timestamp &
                    \texttt{DATE-RED} & \texttt{obs.param; time.epoch} & Timestamp
                    for the processing step being performed.\\ \addlinespace kind &
                    \texttt{N/D} & \texttt{obs.param; meta.code} & Type of
                    processing applied to source data; comes from a controlled
                    vocabulary: \texttt{un\-proc\-essed},
                    \texttt{noise\-Weight\-ed\-Av\-er\-age},
                    \texttt{non\-Weight\-ed\-Av\-er\-age}.\\ \addlinespace
                    software\-Package & \texttt{N/D} & \texttt{meta.software;
                    meta.id} & Software package used for data processing; should
                    come from a controlled vocabulary: \texttt{CLASS},
                    \texttt{AIPS}, \texttt{AIPS++}, \texttt{CASA}, \texttt{MOPSIC},
                    \texttt{GILDAS}, \texttt{MIRA}, \texttt{MIR}, \texttt{other}.
                    In the case of \texttt{other}, the actual package that was used
                    should be added as a parameter, with parameter.name as
                    \texttt{software\-Package} and the parameter.value as the
                    package name. \\ \addlinespace parameter[n].name & \texttt{N/D} &
                    \texttt{obs.param; meta.code} & Additional processing parameter
                    name, whose value will be in parameter.value; eventually, we
                    will have a controlled list of possible parameter.name
                    values.\\ \addlinespace parameter[n].kind & \texttt{N/D} &
                    \texttt{obs.param; meta.code} & From a controlled vocabulary:
                    \texttt{integer}, \texttt{float}, \texttt{string}, et cetera.
                    At least all
                    of FITS data types should be present.\\ \addlinespace
                    parameter[n].value & \texttt{N/D} &
                    \texttt{obs.param}\footnote{The final UCD to mark
                    parameter[n].value will be calculated when writing the VOTable,
                    as it depends on parameter.kind; it will be \texttt{obs.param;
                    meta.number} most of the time, but it could be
                    \texttt{obs.param; meta.name} or \texttt{obs.param; meta.code},
                    depending on the context.} & Value for the parameter indicated
                    by parameter.name.\\ \addlinespace
            \end{smallertabular}
            \label{tabProvenanceProcessingStep}
            \end{minipage}
            \end{table*}

            \begin{table*}
            \begin{minipage}{\linewidth}
            \caption[Calibration metadata]
            {
                Calibration metadata\footnote{It is mandatory that at least one
                \texttt{[parameter.name, parameter.kind, parameter.value]}
                triplet appears, with \texttt{fluxScale} as parameter.name, and
                one of \texttt{antennaTemperature},
                \texttt{mbBrightnessTemperature}, or \texttt{S\_nu} as the
                parameter.value, with a parameter.kind of \texttt{string}.}.
            }
            \begin{smallertabular}{p{2.4cm}p{1.5cm}p{3.8cm}p{7.9cm}}
                    & \textbf{FITS} & & \\ \textbf{Attribute} & \textbf{Keyword} &
                    \textbf{UCD} & \textbf{Description}\\ \midrule \\[-4pt] timestamp &
                    \texttt{DATE-RED} & \texttt{obs.param; time.epoch} & Timestamp
                    for the calibration step being performed.\\ \addlinespace
                    parameter.name & \texttt{N/D} & \texttt{obs.calib;
                    obs.param; meta.id} & Keyword defining the parameter that we
                    will characterise with the remaining attributes.\\
                    \addlinespace parameter.kind & \texttt{N/D} &
                    \texttt{obs.calib; obs.param; meta.code} & Type of calibration
                    parameter used, from a controlled vocabulary:
                    \texttt{additive}, \texttt{fac\-tor}, \texttt{pol\-y\-no\-mi\-al},
                    \texttt{ex\-po\-nen\-tial}, \texttt{log\-a\-rith\-mic}.\\
                    \addlinespace parameter.value & \texttt{N/D} &
                    \texttt{obs.calib; obs.param; meta.number} & Value for the main
                    calibration parameter, where parameter.kind is not
                    \texttt{polynomial}.\\ \addlinespace parameter.sigma &
                    \texttt{N/D} & \texttt{obs.calib; obs.param; meta.number} &
                    Value of sigma, for \texttt{exponential} calibrations.\\
                    \addlinespace parameter.calCoeff.[n] & \texttt{N/D} &
                    \texttt{obs.calib; obs.param; meta.number} & $\mathrm{n^{th}}$
                    degree coefficient for a \texttt{polynomial} calibration
                    parameter; polynomial degree is derived from the maximum n.\\
                    \addlinespace
            \end{smallertabular}
            \label{tabProvenanceCalibration}
            \end{minipage}
            \end{table*}

        
    
    \section{Packaging of observations} 
    \label{sec:packaging_the_vopack}
        
        Single datasets in the VO are typically delivered in one
        of three forms\footnote{Other formats,
        such as JPEG previews (possibly including Astronomical
        Visualization Metadata, 
        \url{http://www.ivoa.net/Documents/latest/AOIMetadata.html}), 
        or CSV values, are technically possible.}: 
        VOTable tabular data, FITS tabular
        data, or FITS observation data (i.e., images, spectra,
        data cubes) with a VOTable description.
        
        However, many datasets need the delivery of more than
        one data product: for instance, OTF with heterodyne
        instruments can produce many different spectra in
        individual scans, and the final reprocessed data cube,
        and users might want to access all of them at the
        same time. Having an organisational unit that binds
        together those different data types, and keeps information
        on the Characterisation of them, their provenance, etc.,
        enabling offline evaluation and selection.
        
        The Packaging class is used to specify how data from
        different sources will be presented together. For
        instance, if we wanted to retrieve data belonging to
        a particular survey subset, a VO system could
        reply with a Multi-Beam FITS file containing all the
        scans and sub-scans that conform the On-Off patterns for
        a single issued observation, or with a \texttt{.zip}
        file with all the FITS files belonging to the survey,
        or with just a single On-Off pair, et cetera.
        
         An instance of the Packaging class describes the
        contents of the data retrieved, in terms of project
        organisation, and of the particular files being
        actually delivered.
        
         As there is no Packaging class defined at
        the VO level, we have resorted to other packaging
        description mechanisms available in other archiving
        tools.
        
         We believe a standard VO packaging scheme is
        need\-ed in order to facilitate distribution of datasets. The
        packaging scheme that we propose, called VOPack ---Virtual
        Observatory Package; \texttt{.vopack} file extension--- is a
        “tarred and gzipped” folder (\texttt{.tgz} file) with an XML
        content descriptor, which acts as VO-aware manifest of
        contained assets.
        
         In particular, a VOPack describes the 
        relationships between the grouped units (VOTable, FITS, and/or
        ancillary files, or included VOPacks), 
        and also provides CharDM metadata for the whole VOPack, 
        which allows to picture the package in observational
        parameter space. 

        The VOPack, then, is a way of distributing VO-compliant
        content, in a way that makes it easy to reuse and point
        to existing content, either remote or locally.
            
\begin{figure}[tb]
        \begin{center}
            \includegraphics[scale=0.5]
            {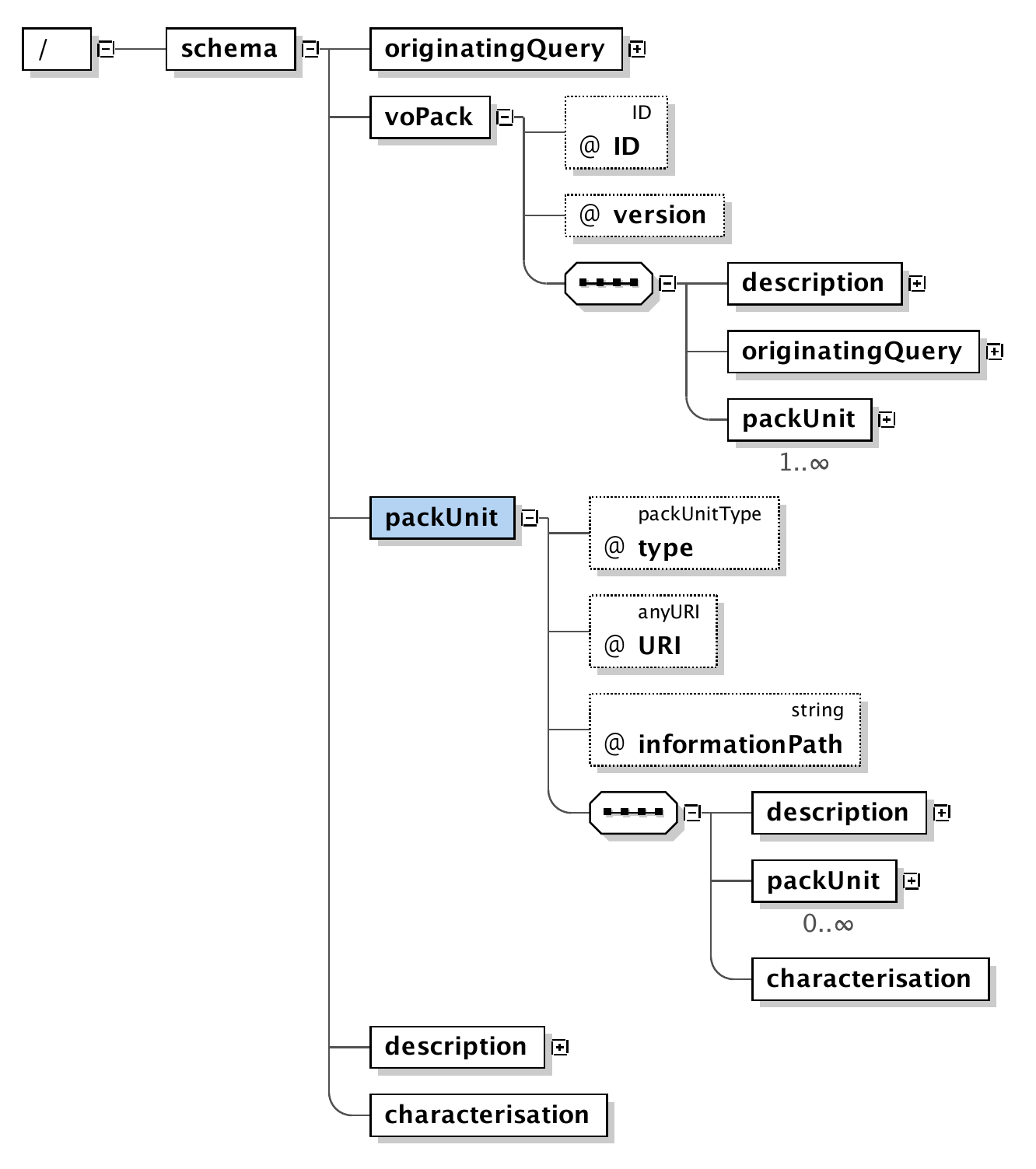}
        \end{center}
        \caption[VOPack structure]
        {
            VOPack structure. Diagram generated by
            \textbf{Oxygen} from the XML schema.
        }
        \label{figVOPackStructure}
        \end{figure}

        VOPacks consist of a compressed file that contains at
        least a \texttt{voPack.xml} file ---following the VOPack
        XML Schema--- that describes all the additional content
        of the compressed VOPack, and their relationships
        between them. Figure~\ref{figVOPackStructure} shows the
        structure of a VOPack from the VOPack schema.
        
        In that diagram, the \texttt{voPack} element is the
        root for the XML document. It includes a description,
        the originating query, and one or more
        \texttt{packUnits}, which actually point to the
        information being retrieved. The
        \texttt{o\-rig\-i\-na\-ting\-Que\-ry}
        element contains the string
        with the URI that allows the retrieval of the voPack.
        Additional \texttt{characterisation} elements,
        following the Characterisation schema, can be used to
        further specify properties on the data being delivered
        with the VOPack.
        
        The \texttt{packUnit} corresponds to a single piece of
        data, or to another \texttt{packUnit}s, in case of more
        structured data. The depth of inclusion is arbitrary.
        
        \texttt{packUnit}s have a \texttt{type} attribute that
        can be one of: \texttt{votable}, \texttt{fits},
        \texttt{vopack}, \texttt{compressedFolder},
        \texttt{folder}, \texttt{otherXML},
        \texttt{otherNonXML}. Table~\ref{packUnitType}
        specifies the meaning of this attribute.
        
        \begin{table*}
            \caption[Valid \texttt{packUnit} attribute values]{
                Meaning of the different valid values for
                attributes of the
                \texttt{packUnit} data type.
            }
            \begin{smalltabular}{rp{13.1cm}}
                \textbf{packUnit} & \textbf{description}
                \\ \midrule 
                
                 \texttt{votable} & The packed unit is a
                VOTable. \\[2pt]
                
                 \texttt{fits} & The packed unit is a FITS
                file. \\[2pt]
                
                 \texttt{vopack} & The packed unit is itself a
                VO pack. The characterisation of the
                referencing VO pack encompasses all packed
                units, while each particular one will have its
                own, \emph{narrower} characterisation.
                \\[2pt]
                
                 \texttt{compressedFolder} & The referenced
                packed unit is a compressed directory.
                \\[2pt]
                
                 \texttt{folder} & The referenced packed unit
                is a directory in the same file-system as the
                referencing VOPack. \\[2pt]
                
                 \texttt{otherXML} & An XML representation,
                other than a VOTable, is used. \\[2pt]
                
                 \texttt{otherNonXML} & A non-XML
                representation, also different from a FITS
                file, is used. This kind of representation
                should be avoided, but would be useful for
                packing instrument specific raw data formats,
                which are correctly characterised.
                \\[2pt] \end{smalltabular}
                \label{packUnitType}
        \end{table*}
        
        For the \texttt{vopack}, \texttt{folder} and
        \texttt{compressedFolder} values, a new
        \texttt{vopack.xml} file has to be provided for their
        description. This allows for meta-packaging of
        ready-made VOPacks.
        
         For the first three types, the
        \texttt{informationPath} attribute gives an XPath to
        the actual data being pointed, just in case the
        \texttt{packUnit} contains several tables, and not all
        of them are to be considered. In the case of FITS
        files, the informationPath looks XPath-like, but points
        to the HDU or Image holding the data.
        
        The VOPack XML Schema has been inspired by the concepts
        of Digital Items, Digital Item Containers, and Digital
        Item Components from MPEG-21~\citep{Bormans:fk}, with the
        aim of being able to reuse the CharDM for direct package
        and sub-package inspection and selection.
        
        The VOPack concept is orthogonal to the rest of RADAMS'
        classes, and can be used independently.
        
    
    \section{RADAMS Uptake and Implementations} 
    \label{sec:implementations_of_the_radams}
        
        The RADAMS has been implemented, so far, in two radio
        astronomical archives: the TAPAS archive for the IRAM~30m 
        radio telescope\urlnote{https://tapas.iram.es/tapas/}, 
        and the DSS-63\urlnote{http://sdc.cab.inta-csic.es/robledo/}.
        
        TAPAS is a complete archive implementation for the IRAM~30m,
        with special emphasis in keeping observation metadata as
        modelled by the RADAMS, and the capability of storing (and
        delivering) data for large programs in the future. It covers
        both heterodyne and bolometer observations for both
        single-pixel and multi-pixel instruments, and has shown its
        modularity by the recent incorporation of the EMIR
        heterodyne
        backends\urlnote{http://iram.fr/IRAMFR/ARN/sep09/node4.html},
        substituting the old ABBA ones.
        
        \begin{figure*}[tbp]
            \centering
                \includegraphics[totalheight=0.9\textheight]
                {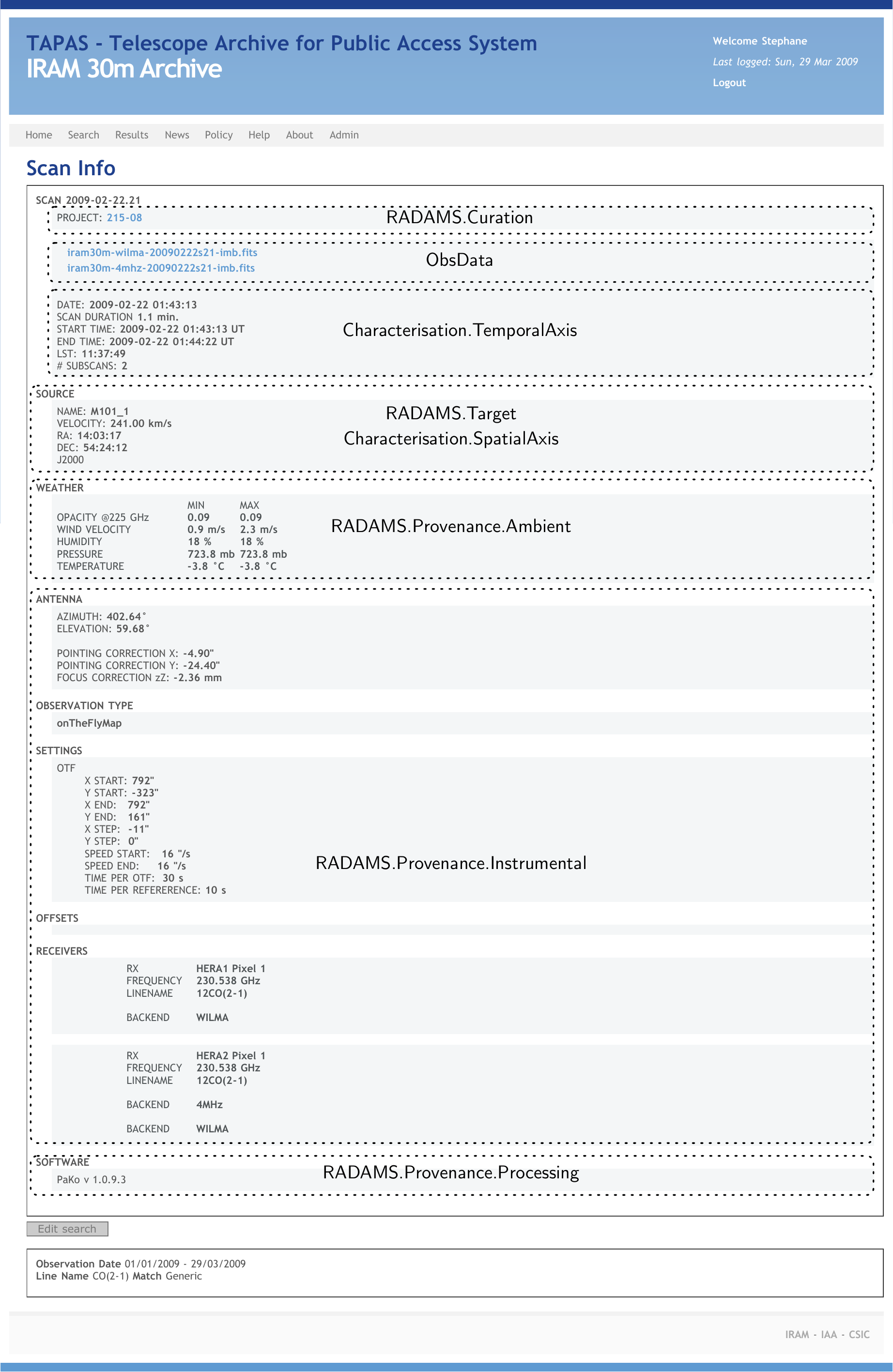}
            \caption{
                Details for a scan belonging to an OTF map,
                indicating the relationship of each metadata element
                with RADAMS submodels.
            }
            \label{fig:fig_TAPAS_scanDetailsOTF}
        \end{figure*}
        
        Figure~\ref{fig:fig_TAPAS_scanDetailsOTF} shows TAPAS output
        for a given observation, with the relationship of the
        different sections to the RADAMS highlighted.
        
         The DSS-63 archive contains the observations in K-band (18
        to 26 GHz) of the DSS-63 antenna taken during 2006. The
        system is prepared for future extensions to other ranges, in
        particular the Q-band (40 to 50 GHz) and Ka-band (32 GHz),
        and forms part of the work developed through a collaboration
        between the AMIGA group (IAA-CSIC) and CAB (INTA-CSIC) for
        the integration of radio astronomical archives in the
        Virtual Observatory.
        
    
    \section{Conclusions \& Future Work} 
    \label{sec:conclusions_future_work}
        
        We have performed a review of the VO enhances 
        astronomer's interaction with astronomical ar\-chives, 
        and how data in
        those archives become interoperable thanks to a unified
        description and semantics, and a shared data model.

         The IVOA has proposed several data models pertaining to
        astronomical observations, specially the description of the
        spatial, temporal, and energy parameters defining observations.
        However, there are no complete instantiation of the complete
		IVOA data 
		model for observations,
        and many radio astronomical specifics are not 
        taken into account in existing models, hence the need for the 
        RADAMS.
        
         Using IVOA principles drafted by the IVOA 
        Data Modelling WG~\citep{2005dmo..rept.....M}, and the
        data model for astronomical data 
        Characterisation~\citep{2008dmadcrept.....L}, we have developed 
        a data model which includes
        Provenance, Policy, Curation, and Packaging classes.
        
         Those added classes are completely modular, both in their
        expression in database form as in their XML serialisation,
        as they use their own UTypes, and as such they can be safely
        ignored by applications not able to understand such
        metadata. Radio astronomical specificities are captured in
        the Provenance class, and the rest of the classes remain
        unchanged. In that way, the RADAMS does not need to be an IVOA standard,
        but can be used as a guide on how to
        provide the missing observation information from IVOA's
        ObsDM framework for radio astronomy, but also for additional bands
        outside of the optical domain.
        
         The RADAMS has been validated by being used as the basis
        for the development of two radio astronomical archives,
        those for the DSS-63 and IRAM~30m antennas. Those archives
        back-up the feasibility of using IVOA data models as the
        basis for the development of new astronomical archives,
		helping with their interoperability.
        
         These archives have needed a complete set of metadata
        for their XML serialisations, consisting of target FITS keywords
        for the Data-filler, and Unified Content Descriptors (UCDs)
        for all attributes. Many UCDs have been built by means of
        juxtaposition of existing ones, so that the resulting UCDs
        are more specific than their individual atoms.
        A few UCDs have been proposed for
        addition to the UCD1+ vocabulary.
        
        We have presented the different modules making up the RADAMS for 
        approval to the IVOA Data Modelling Working Group, in order to 
		provide a reference implementation of an IVOA-based observational 
		data model for radio astronomy.


         We plan to further generalise the RADAMS with support for 
        radio interferometry, and time-series data (i.e., for
        pulsar observations).
        
         Finally, we intend to formulate the Provenance 
        relationships in terms of the Open Provenance Model
        \citep[OPM;][]{Moreau:2010kx}, so that OPM based tools
        can be used to derive attributions, lineage, and provide
        quality metrics.
        
         The SQL source code, and the JSON file describing the
        RADAMS meta-model are available from the authors on request.
        
        
    \begin{acknowledgements}
        The authors acknowledge support from Spanish DGI Grants
        AYA 2005-07516-C02-01, AYA~{}2005-07516-C02-02,
        AYA2008-06181-C02, and Junta
        de Andaluc\'{\i}a grant P08-FQM-4205-PEX. They also acknowledge
        support from the Spanish Virtual Observatory
        (through grants AYA\-2005-24102-E and AYA~{}2008-02156) for useful
        discussions and tra\-vel support for IVOA InterOp meetings.
        They are also 
        grateful for input provided by Jos\'e Francisco
        G\'omez (IAA), Ra\'ul Guti\'errez (CAB), \'Oscar Morata
        (CAB), Carlos Rodrigo-Blan\-co (CAB), and Tom Kuiper
        (JPL) which helped making RA\-DAMS better.
    \end{acknowledgements}
    
    
    \bibliography{bibliography} 
    
\end{document}